\documentclass[11pt]{article}
\usepackage{latexsym,amssymb,amsmath,times}
\usepackage{enumitem}
\usepackage{float}

\usepackage{hyperref}
\usepackage[square,sort&compress,comma,numbers]{natbib}

\bibliography{myrefs}

\makeatletter

\@addtoreset{equation}{section} \makeatother

\input amssym.def
\input amssym.tex

\newcommand{\half}{{{\textstyle\frac{1}{2}}}}
\newcommand{\quarter}{{{\textstyle\frac{1}{4}}}}
\newcommand{\be}{\begin{equation}}
\newcommand{\ee}{\end{equation} }
\newcommand{\beqa}{\begin{eqnarray} }
\newcommand{\eeqa}{\end{eqnarray} }
\newcommand{\ba}{\begin{array}}
\newcommand{\ea}{\end{array}}

\newcommand{\Spin}{\mathbf{Spin}}

\newcommand{\Pin}{\mathbf{Pin}}

\newcommand{\rmd}{{\rm d}}

\newcommand{\mba}{{\mathbf{a}}}
\newcommand{\mbb}{{\mathbf{b}}}
\newcommand{\mbc}{{\mathbf{c}}}
\newcommand{\mbd}{{\mathbf{d}}}

\newcommand{\ODD}{\mathbf{O}(D,D)}
\newcommand{\soDD}{\mathbf{so}(D,D)}

\newcommand{\dSpinD}{{\Spin(1,D{-1})_{\!\it{D}}}}
\newcommand{\SpinD}{{\Spin(1,D{-1})_{\!\it{L}}}}
\newcommand{\oSpinD}{{{\Spin}(D{-1},1)_{\!\it{R}}}}
\newcommand{\oPinD}{{{\Pin}(D{-1},1)_{\!\it{R}}}}

\newcommand{\RC}{\cC}
\newcommand{\RF}{\cF}
\newcommand{\Cp}{{C_{+}}{}}
\newcommand{\Cm}{{C_{-}}{}}
\newcommand{\brCp}{{\brC_{+}}{}}
\newcommand{\brCm}{{\brC_{-}}{}}

\newcommand\Tr{{\rm Tr}}

\newcommand\cA{{\cal A}}

\newcommand\cC{{\cal C}}
\newcommand\cD{{\cal D}}

\newcommand\cF{{\cal F}}

\newcommand\cJ{{\cal J}}

\newcommand\cL{{\cal L}}
\newcommand\cM{{\cal M}}
\newcommand\cN{{\cal N}}

\newcommand\cP{{\cal P}}

\newcommand\cT{{\cal T}}

\newcommand\bcP{{\bar{\cP}}}

\newcommand\hcL{{\hat{\cal L}}}

\newcommand\hdelta{\hat{\delta}}

\newcommand\rhop{{\rho^{\prime}}{}}
\newcommand\psip{\psi^{\prime}{}}

\newcommand\NS{{\scriptscriptstyle{\rm NSNS}}}
\newcommand\RR{{\scriptscriptstyle{\rm RR}}}
\newcommand\typeT{{\scriptscriptstyle{\rm Type\,II}}}

\newcommand\dis{\displaystyle}

\newcommand\seceq{\simeq}

\def\tx{\tilde{x}}

\def\bra{\bar{a}}
\def\brb{\bar{b}}
\def\bre{\bar{e}}

\def\breta{\bar{\eta}}
\def\bralpha{\bar{\alpha}}
\def\brbeta{\bar{\beta}}
\def\brgamma{\bar{\gamma}}
\def\brdelta{\bar{\delta}}
\def\brrho{\bar{\rho}}
\def\brpsi{\bar{\psi}}

\def\brm{{\bar{m}}}
\def\brn{{\bar{n}}}
\def\brp{{\bar{p}}}
\def\brq{{\bar{q}}}
\def\brr{{\bar{r}}}
\def\brs{{\bar{s}}}
\def\bromega{{\bar{\omega}}}

\def\brPhi{{{\bar{\Phi}}}}
\def\brDelta{{{\bar{\Delta}}}}

\def\brC{\bar{C}}
\def\brF{\bar{F}}

\def\brL{\bar{L}}

\def\brV{{\bar{V}}}
\def\brP{{\bar{P}}}

\def\brtau{{\bar{\tau}}}

\def\Tw{{T}}

\newcommand{\DO}{\mathbf{\nabla}}

\newcommand{\na}{{\nabla}}
\newcommand{\trd}{{\bigtriangledown}}

\def\Gammao{\Gamma^{\scriptscriptstyle{0}}}
\def\Phio{\Phi^{\scriptscriptstyle{0}}}
\def\DOo{\DO^{\scriptscriptstyle{0}}}
\def\brPhio{\brPhi^{\scriptscriptstyle{0}}}

\def\cDo{\cD^{\scriptscriptstyle{0}}}
\def\Do{D^{\scriptscriptstyle{0}}}
\def\Fo{F^{\scriptscriptstyle{0}}}
\def\brFo{\brF^{\scriptscriptstyle{0}}}
\def\So{S^{\scriptscriptstyle{0}}}

\begin{document}
\begin{titlepage}
\title{
\vskip 2cm
Ramond-Ramond Cohomology and $\ODD$ T-duality}
\author{\sc Imtak Jeon,${}^{\dagger}$  \mbox{~~\,\,}
Kanghoon Lee${}^{\sharp}$ \mbox{~~\,\,}and\mbox{\,\,~~} Jeong-Hyuck Park${}^{\dagger}$}
\date{}
\maketitle \vspace{-1.0cm}
\begin{center}
~~~\\
${}^{\dagger}$Department of Physics, Sogang University,  Seoul 121-742, Korea\\
~\\
${}^{\sharp}$Center for Quantum Spacetime, Sogang University,  Seoul 121-742, Korea\\
~{}\\
{{\texttt{~{{{imtak@sogang.ac.kr\,,~kanghoon@sogang.ac.kr\,,~park@sogang.ac.kr}}}}}}
~~~\\~\\~\\
\end{center}
\begin{abstract}
\vskip0.2cm
\noindent
In the name of  supersymmetric  double field theory,   superstring effective actions  can be reformulated  into simple forms.  They  feature  a pair  of vielbeins corresponding    to the same spacetime metric, and hence   enjoy  double   local Lorentz symmetries.   In a manifestly covariant manner---with regard  to  $\ODD$ T-duality,  diffeomorphism, $B$-field  gauge symmetry  and the  pair of local Lorentz symmetries---we incorporate R-R potentials    into double field theory. We take  them as a single object which is in a 
 bi-fundamental spinorial   representation of the double   Lorentz groups. We   identify     cohomological structure relevant to the  field strength.     \textit{A priori,}  the R-R sector as well as all the fermions are  $\ODD$ singlet.  Yet, gauge fixing the two vielbeins  equal to each other modifies the $\ODD$ transformation  rule to call for  a compensating local Lorentz rotation, such that the R-R potential may turn into  an $\ODD$ spinor and  T-duality can flip  the chirality   exchanging   type IIA and  IIB supergravities.   
\end{abstract}

{\small
\begin{flushleft}
~~
\end{flushleft}}
\thispagestyle{empty}
\end{titlepage}
\newpage
\tableofcontents 
\section{Introduction} 
Double field theory (DFT)~\cite{Hull:2009mi,Hull:2009zb,Hohm:2010jy,Hohm:2010pp} may    reformulate  closed string effective actions \textit{i.e.~}supergravities into simple forms  and    manifest     the $\ODD$ T-duality  which is a genuine stringy feature~\cite{Buscher:1985kb,Buscher:1987sk,Buscher:1987qj,Giveon:1988tt,Meissner:1991zj}.   The manifestation  is achieved  by  doubling the spacetime dimension, from $D$ to ${D+D}$  with coordinates  $x^{\mu}\rightarrow y^{A}=(\tx_{\mu},x^{\nu})$, where the newly added coordinates $\tx_{\mu}$   correspond to the T-dual coordinates for the closed string  winding mode~\cite{Tseytlin:1990nb,Tseytlin:1990va,Siegel:1993xq,Siegel:1993th}.\\  

In particular, as  for the  $\ODD$ covariant description of the Neveu-Schwarz (NS) sector,  DFT uses ${(D+D)}$-dimensional language or tensors,  equipped with an $\ODD$ invariant constant metric,
\be
\cJ_{AB}=\left(\ba{cc}0&1\\1&0\ea\right)\,.
\label{ODDeta}
\ee
Yet,  DFT  is not truly doubled since it is subject to a \textit{section condition} (or ``strong constraint"~\cite{Hohm:2010jy}):  all the fields are required to live on a $D$-dimensional null hyperplane, such that    the $\ODD$ d'Alembertian operator must be trivial acting on arbitrary fields as well as their products, 
\be
\ba{ll}
\partial_{A}\partial^{A}\Phi\seceq 0\,,~~~~&~~~~\partial_{A}\Phi_{1}\partial^{A}\Phi_{2}\seceq0\,.
\ea
\label{constraint}
\ee
The worldsheet origin of this constraint can be traced back to  the closed string  level-matching condition. \\

Further, DFT  unifies   the  diffeomorphism and the $B$-field  gauge symmetry   into   what we may  call `double-gauge symmetry,' as they are generated by the  generalized Lie derivative~\cite{Courant,Siegel:1993th,Gualtieri:2003dx,Grana:2008yw,Hohm:2010pp},
\be
\hcL_{X}\Tw_{A_{1}\cdots A_{n}}:=X^{B}\partial_{B}\Tw_{A_{1}\cdots A_{n}}+\omega_{{\scriptscriptstyle{T\,}}}\partial_{B}X^{B}\Tw_{A_{1}\cdots A_{n}}+\sum_{i=1}^{n}(\partial_{A_{i}}X_{B}-\partial_{B}X_{A_{i}})\Tw_{A_{1}\cdots A_{i-1}}{}^{B}{}_{A_{i+1}\cdots  A_{n}}\,,
\label{tcL}
\ee
where   $\omega_{{\scriptscriptstyle{T\,}}}$ is the weight of $\Tw_{A_{1}\cdots A_{n}}$  and  $X^{A}$ is the double-gauge symmetry  parameter whose  half  components are for  the  $B$-field  gauge symmetry and the other half are  for the diffeomorphism.  Since Eq.(\ref{tcL}) differs from the ordinary Lie derivative, the underlying  differential geometry of DFT   is not Riemannian. Namely, while doubling the spacetime dimension is sufficient to manifest the $\ODD$ structure,   the double-gauge symmetry (\ref{tcL}) calls for  novel mathematical treatments, such as generalized geometry~\cite{Gualtieri:2003dx,Hitchin:2004ut,Hitchin:2010qz,Grana:2008yw,Coimbra:2011nw,Coimbra:2012yy},  Siegel's  formalism~\cite{Siegel:1993xq,Siegel:1993th,Hohm:2010xe}  and our own  approach~\cite{Jeon:2010rw,Jeon:2011cn,Jeon:2011vx,Jeon:2011sq,Jeon:2011kp,VIDEO} (see also \cite{Hohm:2011si} for a similar  analysis,  \cite{Berman:2010is,Berman:2011cg,Coimbra:2011ky,Berman:2011jh} for  $\cM$-theory extensions, and  especially  \cite{West:2011mm,West:2001as,West:2010ev,Rocen:2010bk} for $E_{11}$ approaches\footnote{For    recent   developments related to   DFT we refer to  \cite{Kwak:2010ew,Hohm:2011ex,Berman:2011pe,Copland:2011yh,Thompson:2011uw,Hohm:2011zr,Hohm:2011dv,Albertsson:2011ux,Hohm:2011cp,Kan:2011vg,Aldazabal:2011nj,Geissbuhler:2011mx,GeissbuhlerThesis,Berman:2011kg,Copland:2011wx,Hohm:2011nu,Copland:2012zz,Kan:2012nf,Grana:2012rr,Andriot:2011uh,Andriot:2012wx,Dibitetto:2012rk,Andriot:2012an,Vaisman2012}.}).\\

Through  the  series of papers~\cite{Jeon:2010rw,Jeon:2011cn,Jeon:2011vx,Jeon:2011sq,Jeon:2011kp},  we have developed a stringy differential geometry  which  manifests,   in a  covariant  manner,  all the symmetries of DFT  listed in Table~\ref{TABsymmetry}.  In particular,  we  conceived  a \textit{semi-covariant derivative}   for the NS-NS sector in \cite{Jeon:2010rw,Jeon:2011cn},  extended it to the fermionic   sector~\cite{Jeon:2011vx}, and  managed to reformulate  the ${\cN=1}$ ${D=10}$ supergravity as a  minimal supersymmetric double field theory (SDFT)  to the  full order in fermions~\cite{Jeon:2011sq}.  We have also applied our formalism to  construct a  double field Yang-Mills theory~\cite{Jeon:2011kp}.   

 \begin{table}[H]
\begin{center}
\begin{itemize}
\item $\ODD$ T-duality: \textit{Meta-symmetry}
\item Gauge symmetries 
\begin{enumerate}
\item Double-gauge symmetry: \textit{Generalized Lie derivative}
\begin{itemize}
\item Diffeomorphism
\item $B$-field gauge symmetry
\end{itemize}
\item A pair of local    Lorentz  symmetries, ${\SpinD\times\oSpinD}$
\end{enumerate}
\end{itemize}
\caption{T-duality and gauge symmetries in DFT.  }
\label{TABsymmetry}
\end{center}
\end{table}

\indent In this paper, within the  geometric  setup~\cite{Jeon:2010rw,Jeon:2011cn,Jeon:2011vx,Jeon:2011sq,Jeon:2011kp},       as a natural next step   toward the construction of  ${\cN=2}$ ${D=10}$ SDFT which should  reformulate  the type IIA and IIB supergravities in a unified manner, \textbf{we   incorporate Ramond-Ramond  (R-R) sector     into double field theory, manifesting the $\ODD$ structure.}  In an apparently   covariant fashion, our formalism respects all the DFT symmetries listed in Table~\ref{TABsymmetry},  including  ${\SpinD\times\oSpinD}$ the double local Lorentz  symmetries.  Further, our formalism   does not require any specific parametrization  of the DFT variables ($V_{A}{}^{p}$, $\brV_{A}{}^{\brp}$, \textit{etc.})  in terms of   the metric $g_{\mu\nu}$ and the Kalb-Ramond $B$-field, and  is independent of    the choice of the $D$-dimensional null hyperplane   for the section condition (\ref{constraint}),  like  $\frac{\partial~~}{\partial x^{\mu}}\seceq 0$ \,or\, $\frac{\partial~~}{\partial \tilde{x}_{\mu}}\seceq 0$.     \\

 Preceding  related works    include the papers   by  Fukuma, Oota and Tanaka~\cite{Fukuma:1999jt}, by  Hassan~\cite{Hassan:1999bv,Hassan:1999mm,Hassan:2000kr}, by Berkovits and Howe~\cite{Berkovits:2001ue},  by Coimbra,  Strickland-Constable and  Waldram~\cite{Coimbra:2011nw,Coimbra:2012yy},  and  by Hohm,  Kwak and Zwiebach~\cite{Hohm:2011zr,Hohm:2011dv}, where  the R-R sector was treated   as an  $\ODD$ spinor~\cite{Fukuma:1999jt,Hassan:1999mm,Hohm:2011zr,Hohm:2011dv}  or as  a $D$-dimensional bi-spinor~\cite{Hassan:1999bv,Hassan:1999mm,Hassan:2000kr,Berkovits:2001ue,Coimbra:2011nw,Coimbra:2012yy}. It was pointed out by   Hassan that the $\ODD$ transformations of the R-R sector in the two approaches are equivalent being  compatible  with the supersymmetry of type IIA/IIB supergravities~\cite{Hassan:1999mm}. However,  while the bi-spinorial R-R field  is ready to   couple naturally to fermions   for supersymmetry (\textit{e.g.~}the `democratic' formalism~\cite{Townsend:1995gp,Green:1996bh,Bergshoeff:2001pv} and   a pure spinor approach~\cite{Benichou:2008it}),     the $\ODD$ spinorial R-R field  appears rather awkward to do so. \\

In this work, we assert to put  the R-R sector \textit{a priori}  in  the bi-fundamental spinorial  representation of  ${\SpinD\times\oSpinD}$, rather than  in the $\ODD$ spinorial  representation.   One crucial novel point  
in  our work  is that, compared to the precedents and contrary to the well-known proposition,   \textbf{the  R-R potential and the NS-R/R-NS fermions are   \textit{a priori} all   $\ODD$ singlet in our covariant  DFT formalism}, such that  the $\ODD$ T-duality does not exchange type IIA and IIB   supergravities!   After  gauge fixing  the double  local    Lorentz  symmetries to be their  diagonal subgroup,  the $\ODD$ transformation rule  gets  modified in order to preserve the gauge choice. Namely, the $\ODD$ T-duality  now   rotates  not only the $\ODD$ vector indices but also  local Lorentz indices: more precisely,  one of the double local Lorentz indices which we choose  below   to be $\oSpinD$ rather than $\SpinD$ without loss of generality.  That is to say, the $\oSpinD$ indices are no longer $\ODD$ singlet after the gauge fixing. In particular,  the R-R potential    and the NS-R fermion can  flip their chiralities,  resulting in     the   exchange of type IIA and IIB supergravities. This   essentially recovers  the results  by Hassan~\cite{Hassan:1999bv,Hassan:1999mm,Hassan:2000kr} (see also  \textit{e.g.~}\cite{Grana:2008yw,Lust:2008zd,Caviezel:2009tu,Blaback:2010sj,Sfetsos:2010uq,Lozano:2011kb,Sfetsos:2011jw}  for related   recent progress). We also show that the diagonal  gauge fixing may turn the R-R potential into   an $\ODD$ spinor verifying  the result by  Hohm,  Kwak and Zwiebach~\cite{Hohm:2011zr,Hohm:2011dv}. \\

A similar mechanism holds for  Dirac fermions  in ordinary quantum field theories  on flat Minkowskian spacetime.  We may gauge the  internal  Lorentz symmetry by introducing a  spin connection which is made of a flat vielbein, and hence corresponds to a pure gauge. Then  Dirac fermions  are singlet for the global spacetime Lorentz symmetry. However, gauge fixing the vielbein to be trivial  breaks the local Lorentz symmetry, and the fermions start to    transform as a spacetime spinor  under the global  Lorentz symmetry.   Another analogous example  is  the  metamorphosis of the spacetime fermion into a worldsheet spinor after a gauge fixing of the kappa-symmetry  in  the Green-Schwarz superstring action.\\

\textbf{The main contents as well as the organization of the present  paper are as follows. }\\

We separate  out the main body  into two parts.   Part I is genuinely `double-field-theoretical' being independent of the parametrization of the DFT variables  in terms of the metric, $g_{\mu\nu}$, and the Kalb-Ramond field, $B_{\mu\nu}$.  Part II       deals with   a specific parametrization and  conveys the  modified  $\ODD$ transformation rule after the diagonal gauge fixing. 

\begin{enumerate}
\item \underline{Part I:  parametrization independent formalism  where  R-R sector is   $\ODD$ singlet.}
\begin{itemize}

\item Section \ref{SECRR} contains our covariant DFT formalism especially for  R-R sector. As for a unifying description of all the R-R potentials,  we consider    a  single bosonic object   which is in a bi-fundamental spinorial   representation of the double  local Lorentz groups. In particular,   we   construct 
 \textit{a pair of nilpotent differential operators which can act on the R-R potential and define the field strength within  the  DFT  formalism}.

\item In section \ref{SECtypeII}, we spell out the bosonic part of the type II (or  ${\cN=2}$) supersymmetric double field theory Lagrangian which corresponds to the DFT reformulation of the type II democratic supergravity~\cite{Bergshoeff:2001pv}. We derive the equations of motion  and discuss the self-duality of the R-R field strength.

\end{itemize}

\item  \underline{Part II: specific parametrization and  gauge fixing where  R-R sector is   $\ODD$ non-singlet.}
\begin{itemize}
\item In section \ref{SECpara},  we  parametrize the  covariant DFT variables in terms of a pair of $D$-dimensional  vielbeins and a Kalb-Ramond $B$-field.  We consider   a diagonal gauge fixing of the double  local Lorentz symmetries, $\SpinD\times\oSpinD$, by equating  the two vielbeins.   We show that, after the gauge fixing,  the  $\ODD$ transformation rule must be modified to call for     a compensating local Lorentz rotation.  We  verify that  the gauge fixing may turn   the R-R potential  into an $\ODD$ spinor, and  further that T-duality can flip the chiralities    of the R-R sector  and  the NS-R   fermions. This   manifestly  realizes   the exchange of   type IIA and IIB supergravities.
\end{itemize}
\item Section~\ref{SECCON} contains our \underline{conclusion}.

\item  In \underline{Appendix}, we review in a self-contained  manner  the stringy differential geometry of SDFT  developed in \cite{Jeon:2010rw,Jeon:2011cn,Jeon:2011vx,Jeon:2011sq,Jeon:2011kp}.  We set  our conventions,   spell out   all the  $\ODD$ covariant   fundamental   field variables   constituting    type II  SDFT,   summarize  various    fully covariant  quantities with respect to all the symmetries in Table~\ref{TABsymmetry},  and discuss the reduction to ordinary Riemannian geometry. 
\end{enumerate}
\newpage
\section{R-R sector and Cohomology, before gauge fixing\label{SECRR}}
\subsection{R-R sector in SDFT}
The  NS-NS sector  of    SDFT consists of  DFT-dilaton, $d$,  and 
double-vielbeins, $V_{Ap}$, $\brV_{A\brp}$~\cite{Jeon:2011cn,Jeon:2011vx,Jeon:2011sq}.\footnote{For the full field contents of type II SDFT, see Table~\ref{TABfields} in Appendix.} 
For the R-R sector,   we consider    R-R potential, $\cC^{\alpha}{}_{\bralpha}$, which  is -- as the indices indicate --  in the  \textit{bi-fundamental spinorial  representation of     the double local Lorentz group},  $\SpinD\times\oSpinD$ (\textit{cf.~}\cite{Coimbra:2011nw,Hassan:1999mm,Hassan:2000kr}), while being double-gauge and $\ODD$ singlet.  More precisely,   `fundamental' with respect to   $\SpinD$,  and  `anti-fundamental' with respect to   $\oSpinD$,  especially when the spinorial  indices are suppressed.  However, the indices can be freely  lowered or raised by the symmetric  charge conjugation matrices, $C_{+}{}_{\alpha\beta}$,   $\brC_{+\bralpha\brbeta}$ (\ref{CcM}), and the distinction of being fundamental and anti-fundamental is unimportant.\\

The R-R potential   must satisfy a `chirality' condition,
\be
\gamma^{(D+1)}\cC\brgamma^{(D+1)}=\pm\,\cC\,.
\label{RRDS}
\ee
Hereafter,   the upper sign is for type IIA and the lower  sign is for   type IIB.\\

As for the  differential operators of the R-R sector, we present  a pair of   covariant derivatives  which can be applied to any  $\SpinD\times\oSpinD$ bi-fundamental field, $\cT^{\alpha}{}_{\brbeta}$,
\be
\ba{ll}
\cDo_{+}\cT:=\gamma^{A}\Do_{A}\cT+\gamma^{(D+1)}\Do_{A}\cT\brgamma^{A}\,,~~~~&~~~~
\cDo_{-}\cT:=\gamma^{A}\Do_{A}\cT-\gamma^{(D+1)}\Do_{A}\cT\brgamma^{A}\,.
\ea
\label{DpmS}
\ee
Here the superscript `$0$' indicates that the semi-covariant derivatives assume   the torsionless connection~(\ref{Gammao}).    We stress that, these differential  operators  are  covariant with respect to all the symmetries of DFT listed in Table~\ref{TABsymmetry}. Further, as we show below in section~\ref{SECnil},  they  are \textit{nilpotent},  up to the  section condition  (\ref{constraint}),
\be
\ba{ll}
(\cDo_{+})^{2}\cT\seceq 0\,,~~~~~&~~~~~(\cDo_{-})^{2}\cT\seceq0\,,
\ea
\label{nilpotentS} 
\ee 
and hence, they  define cohomology.   \\

It is worth while  to note
\be
\ba{ll}
\cDo_{\pm}(\gamma^{(D+1)}\cT)=-\gamma^{(D+1)}\cDo_{\mp}\cT\,,~~~~&~~~~
\cDo_{\pm}(\cT\brgamma^{(D+1)})=(\cDo_{\mp}\cT)\brgamma^{(D+1)}\,.
\ea
\label{DgDS}
\ee
~\\

We define the R-R field strength using one of the nilpotent differential operators (\ref{DpmS}), 
\be
\cF:=\cDo_{+}\RC=\gamma^{A}\cDo_{A}\RC+\gamma^{(D+1)}\cDo_{A}\RC\brgamma^{A}\,.
\label{cFdef}
\ee
This quantity  is also $\SpinD\times\oSpinD$ bi-fundamental, and from  (\ref{DgDS}) it  carries   the opposite chirality,
\be
\cF=\mp\gamma^{(D+1)}\cF\brgamma^{(D+1)}\,.
\label{Fchirality}
\ee
Further, thanks to the nilpotency (\ref{nilpotentS}),   the R-R gauge symmetry is simply realized by the same differential operator,
\be
\ba{lll}
\delta\RC=\cDo_{+}\Delta~~&~
\Longrightarrow~&~~\delta\RF=\cDo_{+}(\delta\cC)=(\cDo_{+})^{2}\Delta\seceq 0\,,
\ea
\ee
where $\Delta^{\alpha}{}_{\bralpha}$ is  an arbitrary  gauge  parameter which is  in the $\SpinD\times\oSpinD$ bi-fundamental representation and  satisfies  the  same chirality condition as the field strength,
\be
\Delta=\mp\gamma^{(D+1)}\Delta\brgamma^{(D+1)}\,.
\ee
~\\

The  R-R sector Lagrangian, $\cL_{\RR}$, in type II SDFT assumes   a compact  form:
\be
\cL_{\RR}=-\half e^{-2d}\cF^{\alpha\bralpha}\cF_{\alpha\bralpha}\,.
\label{RRL}
\ee
Under  arbitrary variations of all the elementary bosonic DFT  fields, $V_{Ap}$, $\brV_{A\brp}$, $d$ and $\cC^{\alpha}{}_{\bralpha}$,  the R-R sector Lagrangian transforms, up to total derivatives ($\,\cong\,$),  from (\ref{useful0}), (\ref{useful1}), (\ref{useful2}), (\ref{useful3}), as
\be
\ba{l}
\delta\cL_{\RR}=\delta(-\half e^{-2d}\cF^{\alpha\bralpha}\cF_{\alpha\bralpha})\\
~\cong e^{-2d}\left(\delta\cC -\cC\delta d 
+\quarter V^{A}{}_{p}\delta V_{Aq}\gamma^{pq}\cC+\half\brV^{A}{}_{\brp}\delta V_{Aq}\gamma^{(D+1)}\gamma^{q}\cC\brgamma^{\brp}-\quarter\brV^{A}{}_{\brp}\delta\brV_{A\brq}\cC\brgamma^{\brp\brq}
\right)^{\alpha\bralpha}\left(\cDo_{-}\cF\right)_{\alpha\bralpha}\\
{}~~\quad-\half e^{-2d}\,\brV^{A}{}_{\brq}\delta V_{Ap}(\gamma^{p}\gamma^{(D+1)}\cF\brgamma^{\brq})^{\alpha\bralpha\,}\cF_{\alpha\bralpha}\,.
\ea
\label{RRvar}
\ee
It is remarkable that   both  the chiral part of $\delta V_{Ap}$ and the anti-chiral part of $\delta\brV_{A\brp}$, as well as the variation of the DFT-dilaton,   commonly lead to nothing but the equation of motion for the R-R potential,\footnote{Similar phenomena  occur in $\cN=1$ SDFT with   fermions,  see  Eq.(33) of Ref.\cite{Jeon:2011sq}. The observation for the variation of the DFT-dilaton  was also   made by  David Geissbuhler~\cite{privateG}.}
\be
\cDo_{-}\cF=\cDo_{-}\cDo_{+}\cC=0\,.
\label{EOMRR}
\ee
Nevertheless,  as we continue to discuss below in section~\ref{SECtypeII},   an additional  self-duality  relation~(\ref{SD})   needs to be imposed  on  the R-R field strength.\\

\subsection{Cohomology\label{SECnil}}
In this subsection,  we show  the nilpotency (\ref{nilpotentS}) of the differential operators, $\cDo_{\pm}$,  which are defined to act on an arbitrary  $\SpinD\times\oSpinD$ bi-fundamental spinorial field. \\

To sketch our proof, we set  with (\ref{threeF}) some notations,
\be
\ba{l}
\Phio_{A}=\quarter\Phio_{Apq}\gamma^{pq}=\quarter(V^{B}{}_{p}\partial_{A}V_{Bq}+\Gammao_{Apq})\gamma^{pq}
\,,\\
\brPhio_{A}=\quarter\brPhio_{A\brp\brq}\brgamma^{\brp\brq}=\quarter(\brV^{B}{}_{\brp}\partial_{A}\brV_{B\brq}+\Gammao_{A\brp\brq})\brgamma^{\brp\brq}\,,\\
\Fo_{AB}=\partial_{A}\Phio_{B}-\partial_{B}\Phio_{A}+\left[\Phio_{A},\Phio_{B}\right]=\quarter \Fo_{ABpq}\gamma^{pq}\,,\\
\brFo_{AB}=\partial_{A}\brPhio_{B}-\partial_{B}\brPhio_{A}+\left[\brPhio_{A},\brPhio_{B}\right]=\quarter \brFo_{AB\brp\brq}\brgamma^{\brp\brq}\,.
\ea
\ee
Also, if no confusion arises,   we may  convert  an $\ODD$ vector index   either to a $\SpinD$  or to a  $\,\oSpinD$  vector index  {via}  contraction   with the double-vielbein, $V^{A}{}_{p}\,$ or  $\,\brV^{A}{}_{\brp}$ respectively,  such as   (\ref{Dpbrp}), (\ref{Spbrp}).\\

Without loss of generality, for simplicity we consider  an arbitrary bi-fundamental spinor, $\cT^{\alpha}{}_{\bralpha}$, which has zero weight. We begin with the expression, 
\be
(\cDo_{\pm})^{2}\cT=\cDo{}^{A}\cDo_{A}\cT+\half\gamma^{AB}\left[\cDo_{A},\cDo_{B}\right]\cT-\half\left[\cDo_{A},\cDo_{B}\right]\cT\brgamma^{AB}\mp\gamma^{(D+1)}\gamma^{A}\left[\cDo_{A},\cDo_{B}\right]\cT\brgamma^{B}\,,
\label{beginEQ}
\ee
into which we need to substitute
\be
\ba{ll}
{}\left[\cDo_{A},\cDo_{B}\right]\cT&=-\Gammao{}^{C}{}_{AB}\cDo_{C}\cT+\Fo_{AB}\cT-\cT\brFo_{AB}\\
{}&=-\Gammao{}^{C}{}_{AB}\partial_{C}\cT+(\Fo_{AB}-\Gammao{}^{C}{}_{AB}\Phio_{C})\cT-
\cT(\brFo_{AB}-\Gammao{}^{C}{}_{AB}\brPhio_{C})\,,
\ea
\ee
and 
\be
\ba{ll}
\cDo{}^{A}\cDo_{A}\cT&\seceq(\partial^{A}\Phio_{A}-\Phio{}^{A}\Phio_{A}+\Gammao_{A}{}^{AB}\Phio_{B})\cT-\cT(\partial^{A}\brPhio_{A}+\brPhio{}^{A}\brPhio_{A}+\Gammao_{A}{}^{AB}\brPhio_{B})\\
{}&~~~~~~+2\Phio{}^{A}\cT\brPhio_{A}+2\Phio{}^{A}\cDo_{A}\cT-2\cDo_{A}\cT\brPhio{}^{A}\,.
\ea
\ee
~\\
The first three terms on the right hand side of the equality  in (\ref{beginEQ}) then give
\be
\ba{l}
\cDo{}^{A}\cDo_{A}\cT+\half\gamma^{AB}\left[\cDo_{A},\cDo_{B}\right]\cT-\half\left[\cDo_{A},\cDo_{B}\right]\cT\brgamma^{AB}\\
\quad\quad\quad\quad~~~~~\seceq \left[\partial_{A}\Phio{}^{A}+\Phio_{A}\Phio{}^{A}+\half\gamma^{AB}\Fo_{AB}+\left(\Gammao{}^{B}{}_{BA}-\half\Gammao_{Apq}\gamma^{pq}\right)\Phio{}^{A}\right]\cT\\
\quad\quad\quad\quad~~~~~~~~~~-\cT\left[\partial_{A}\brPhio{}^{A}-\brPhio_{A}\brPhio{}^{A}-\half\brFo_{AB}\brgamma^{AB}+\brPhio_{A}\left(\Gammao{}_{B}{}^{BA}+\half\Gammao{}^{A}{}_{\brp\brq}\brgamma^{\brp\brq}\right)\right]\\
~~~~\quad\quad\quad\quad~~~~~~-2\Phio{}^{A}\cT\brPhio_{A}-\half\left(\Fo_{\brp\brq}-\Gammao{}^{C}{}_{\brp\brq}\Phio_{C}\right)\cT\brgamma^{\brp\brq}-\half\gamma^{pq}\cT\left(\brFo_{pq}-\Gammao{}^{C}{}_{pq}\brPhio_{C}\right).
\ea
\label{DDT3}
\ee
Due to the following identities which can be shown by brute force computation,   
\be
\ba{l}
\partial_{A}\Phio{}^{A}+\Phio_{A}\Phio{}^{A}+\half\gamma^{AB}\Fo_{AB}+\left(\Gammao{}^{B}{}_{BA}-\half\Gammao_{Apq}\gamma^{pq}\right)\Phio{}^{A}\seceq-\quarter\So_{ABCD}P^{AC}P^{BD}\,,\\
\partial_{A}\brPhio{}^{A}-\brPhio_{A}\brPhio{}^{A}-\half\brFo_{AB}\brgamma^{AB}+\brPhio_{A}\left(\Gammao{}_{B}{}^{BA}+\half\Gammao{}^{A}{}_{\brp\brq}\brgamma^{\brp\brq}\right)\seceq+\quarter\So_{ABCD}\brP^{AC}\brP^{BD}\,,
\ea
\ee
the first two lines of the right hand side of (\ref{DDT3}) get simplified, and in fact from  (\ref{Sidequiv}), they vanish,
\be
\ba{l}
\left[\partial_{A}\Phio{}^{A}+\Phio_{A}\Phio{}^{A}+\half\gamma^{AB}\Fo_{AB}+\left(\Gammao{}^{B}{}_{BA}-\half\Gammao_{Apq}\gamma^{pq}\right)\Phio{}^{A}\right]\cT\\
\quad~~~~~~-\cT\left[\partial_{A}\brPhio{}^{A}-\brPhio_{A}\brPhio{}^{A}-\half\brFo_{AB}\brgamma^{AB}+\brPhio_{A}\left(\Gammao{}_{B}{}^{BA}+\half\Gammao{}^{A}{}_{\brp\brq}\brgamma^{\brp\brq}\right)\right]\\
\quad\seceq-\quarter(P^{AC}P^{BD}+\brP^{AC}\brP^{BD})\So_{ABCD}\cT\\
\quad\seceq 0\,.
\ea
\ee
Further, from (\ref{Bianchi}),  we have
\be
\Fo_{\brp\brq pq}+\brFo_{pq\brp\brq}=2\So_{pq\brp\brq}+\Gammao{}^{C}{}_{pq}\Gammao{}_{C\brp\brq} 
\seceq\Gammao{}^{C}{}_{pq}\Gammao{}_{C\brp\brq} \,.
\label{FFG}
\ee
Consequently, with  (\ref{FFG}), the remaining terms of the right hand side of (\ref{DDT3}) vanish too,
\be
\ba{l}
\Phio{}^{A}\cT\brPhio_{A}+\quarter\left(\Fo_{\brp\brq}-\Gammao{}^{C}{}_{\brp\brq}\Phio_{C}\right)\cT\brgamma^{\brp\brq}+\quarter\gamma^{pq}\cT\left(\brFo_{pq}-\Gammao{}^{C}{}_{pq}\brPhio_{C}\right)\\
\quad\seceq\textstyle{\frac{1}{16}}\left(\Fo_{\brp\brq pq}+\brFo_{pq\brp\brq}+\Phio{}^{A}{}_{pq}\brPhio_{A\brp\brq}
-\Gammao{}^{A}{}_{pq}\brPhio_{A\brp\brq}-\Phio{}^{A}{}_{pq}\Gammao{}_{A\brp\brq}\right)\gamma^{pq}\cT\brgamma^{\brp\brq}\\
\quad\seceq\textstyle{\frac{1}{16}}\left(\Fo_{\brp\brq pq}+\brFo_{pq\brp\brq}-
\Gammao{}^{A}{}_{pq}\Gammao{}_{A\brp\brq}\right)\gamma^{pq}\cT\brgamma^{\brp\brq}\\
\quad\seceq\textstyle{\frac{1}{8}}\So_{pq\brp\brq}\gamma^{pq}\cT\brgamma^{\brp\brq}\\
\quad\seceq 0\,.
\ea
\ee
Thus, (\ref{DDT3}) vanishes completely. \\

Finally, in order to see  the last term  in (\ref{beginEQ}) vanish, we need identities  coming  from  (\ref{RFF}), (\ref{Sdef}), 
\be
\ba{ll}
\Fo_{p\brq rs}-\Gammao{}^{C}{}_{p\brq}\Gammao{}_{Crs}=2\So_{p\brq rs}\,,~~~~&~~~~
\brFo_{p\brq\brr\brs}-\Gammao{}^{C}{}_{p\brq}\Gammao{}_{C\brr\brs}=2\So_{p\brq\brr\brs}\,,
\ea
\ee
from  (\ref{USEFULv}),
\be
\ba{ll}
\Gammao{}^{C}{}_{p\brq}\Phio_{Crs}\seceq \Gammao{}^{C}{}_{p\brq}\Gammao{}_{Crs}\,,~~~~&~~~~
\Gammao{}^{C}{}_{p\brq}\brPhio_{C\brr\brs}\seceq \Gammao{}^{C}{}_{p\brq}\Gammao{}_{C\brr\brs}\,,
\ea
\ee
and from the Bianchi identity (\ref{Bianchi}),
\be
\ba{ll}
\So_{p\brq rs}\gamma^{p}\gamma^{rs}=2P^{AB}\So_{A\brq Bs}\gamma^{s}\,,~~~~&~~~~
\So_{p\brq\brr\brs}\brgamma^{\brr\brs}\brgamma^{\brq}=2\brP^{AB}\So_{ApB\brs}\brgamma^{\brs}\,.
\ea
\ee
Thanks to  these identities,  the last term in  (\ref{beginEQ}) gets simplified,  and eventually,   with  the second  identity in (\ref{Bianchi}), it vanishes,
\be
\ba{ll}
\gamma^{A}\left[\cDo_{A},\cDo_{B}\right]\cT\brgamma^{B}\!\!&\seceq
\quarter\gamma^{p}\left[\left(\Fo_{p\brq rs}-\Gammao{}^{A}{}_{p\brq}\Gammao_{Ars}\right)\gamma^{rs}\cT
-\cT\gamma^{\brr\brs}\left(\brFo_{p\brq\brr\brs}-\Gammao{}^{A}{}_{p\brq}\Gammao_{A\brr\brs}\right)
\right]\brgamma^{\brq}\\
{}&\seceq\half\gamma^{p}\left[\So_{p\brq rs}\gamma^{rs}\cT
-\cT\gamma^{\brr\brs}\So_{p\brq\brr\brs}\right]\brgamma^{\brq}\\ 
{}&\seceq(P^{AB}-\brP^{AB})\So_{ApB\brq}\gamma^{p}\cT\brgamma^{\brq}\\
{}&\seceq 0\,.
\ea
\ee
This completes our proof of  the nilpotency.  \\
~\\
In a similar fashion, the following operators,
\be
\ba{ll}
\widetilde{\cD}^{\scriptscriptstyle{0}}_{+}\cT:=\gamma^{A}\Do_{A}\cT\brgamma^{(D+1)}+\Do_{A}\cT\brgamma^{A}\,,~~~~&~~~~
\widetilde{\cD}^{\scriptscriptstyle{0}}_{-}\cT:=\gamma^{A}\Do_{A}\cT\brgamma^{(D+1)}-\Do_{A}\cT\brgamma^{A}\,,
\ea
\ee 
can be also shown to be  nilpotent,
\be
\ba{ll}
(\widetilde{\cD}^{\scriptscriptstyle{0}}_{+})^{2}\seceq 0\,,~~~~~&~~~~~(\widetilde{\cD}^{\scriptscriptstyle{0}}_{+})^{2}\seceq 0\,.
\ea
\ee
However, in this work,  our main interest  lies in the R-R potential, $\cC^{\alpha}{}_{\bralpha}$,  which is a bi-fundamental spinor  satisfying the chirality  condition, $\cC =\pm\gamma^{(D+1)}\cC\brgamma^{(D+1)}$ (\ref{RRDS}).  We have then  
\be
\ba{ll}
\widetilde{\cD}^{\scriptscriptstyle{0}}_{+}\cC=\left(\cDo_{\mp}\cC\right)\brgamma^{(D+1)}\,,~~~~~&~~~~~
\widetilde{\cD}^{\scriptscriptstyle{0}}_{-}\cC=\left(\cDo_{\pm}\cC\right)\brgamma^{(D+1)}\,.
\ea
\ee
Therefore, the differential operators  become degenerate. For this reason, in this paper we focus on the operators, $\cDo_{\pm}$ (\ref{DpmS}).\\
~\\

\section{Type II  Democratic  Double Field Theory\label{SECtypeII}}
Combining   the  NS-NS sector DFT Lagrangian~(\ref{NSNSL})~\cite{Jeon:2011cn,Jeon:2011sq} and the R-R sector DFT Lagrangian~(\ref{RRL}),
we are able to spell out  the bosonic part of  type II or ${\cN=2}$ SDFT Lagrangian,  
\be
\cL_{\typeT}=\cL_{\NS}+\cL_{\RR}=e^{-2d}\left[\textstyle{\frac{1}{8}}(P^{AB}P^{CD}-\brP^{AB}\brP^{CD})\So_{ACBD}-
\half\Tr(\cF\bar{\cF})\right]\,,
\label{typeIIL}
\ee
where $\bar{\cF}^{\bralpha}{}_{\alpha}$ denotes the charge conjugation,
\be
\bar{\cF}:=\brC_{+}^{-1}\cF^{T}C_{+}\,,
\label{brcF}
\ee
and the trace is over the $\SpinD$ spinorial index,  such that $\Tr(\cF\bar{\cF})=\cF^{\alpha\bralpha}\cF_{\alpha\bralpha}$.\\

Under  arbitrary variations of all the  bosonic  fields,  from  (\ref{RRvar}), (\ref{Svar}),  (\ref{useful2}), 
the  Lagrangian  transforms, up to total derivatives ($\,\cong\,$), as  
\be
\ba{l}
\delta\cL_{\typeT}\\
~\cong-\quarter e^{-2d}\delta d\,\left(P^{AB}P^{CD}-\brP^{AB}\brP^{CD}\right)\So_{ACBD}\\
~\quad+\half e^{-2d}\brV_{A}{}^{\brq}\delta V^{Ap}\,\left[\So_{p\brq}-
\Tr(\gamma_{p}\gamma^{(D+1)}\cF\brgamma_{\brq}\bar{\cF})\right]\\
~\quad+e^{-2d}\Tr\Big[\!\left(\delta\cC-\cC\delta d
+\quarter V^{A}{}_{p}\delta V_{Aq}\gamma^{pq}\cC+\half\brV^{A}{}_{\brp}\delta V_{Aq}\gamma^{(D+1)}\gamma^{q}\cC\brgamma^{\brp}-\quarter\brV^{A}{}_{\brp}\delta\brV_{A\brq}\cC\brgamma^{\brp\brq}
\right)\overline{\cDo_{-}\cF}\,\Big]\,,
\ea
\label{opf}
\ee
where like (\ref{brcF}), $\,\overline{\cDo_{-}\cF}=\brC_{+}^{-1}(\cDo_{-}
\cF)^{T}C_{+}$.   The \textit{equations of motion} are then  as follows.\\

\begin{itemize}
\item For the DFT-dilaton, $d$, 
\be
\left(P^{AB}P^{CD}-\brP^{AB}\brP^{CD}\right)\So_{ACBD}=0\,.
\label{LNSd}
\ee
Namely the NS-NS Lagrangian vanishes on-shell, $\cL_{\NS}=0$.
\item For the double-vielbein, $V_{A}{}^{p}$, $\brV_{A}{}^{\brp}$, we have the DFT generalization of the Einstein equation, 
\be
\So_{p\brq}-
\Tr(\gamma_{p}\gamma^{(D+1)}\cF\brgamma_{\brq}\bar{\cF})=0\,.
\label{EinsteinDFT}
\ee
\item For the R-R potential, $\cC^{\alpha}{}_{\bralpha}$,  the equation of motion is,  as anticipated in (\ref{EOMRR}),
\be
\cDo_{-}\cF=\cDo_{-}\cDo_{+}\cC=0\,.
\label{EOMRR2}
\ee
However,  the above type II democratic DFT Lagrangian~(\ref{typeIIL}) is supposed to be pseudo~\cite{Bergshoeff:2001pv}:  an additional  self-duality  relation  needs to be imposed  on  the R-R field strength  by hand,
\be
\ba{ll}
\cF=\gamma^{(D+1)}\cF=\mp\cF\brgamma^{(D+1)}\quad:~&~\mbox{Self-Duality\,.}
\ea
\label{SD}
\ee
\end{itemize}
In Eq.(\ref{SD}),  the second equality holds due to  the first one and the chirality~(\ref{Fchirality}).  
From (\ref{DgD}), it is clear that the self-duality (\ref{SD}) ensures the equation of motion (\ref{EOMRR2}) to hold,
\be
\cDo_{-}\cF=\cDo_{-}\left(\gamma^{(D+1)}\cF\right)=-\gamma^{(D+1)}\cDo_{+}\cF=-\gamma^{(D+1)}(\cDo_{+})^{2}\cC\seceq 0\,.
\ee
Further, since $C_{+}\gamma^{(D+1)}$ is anti-symmetric, the self-duality implies that the R-R sector  Lagrangian vanishes too,
\be
\cL_{\RR}=-\half e^{-2d}\Tr(\cF\bar{\cF})=-\half e^{-2d}\Tr(\gamma^{(D+1)}\cF\bar{\cF})=0\,.
\ee
Therefore, with (\ref{LNSd}), the whole   Lagrangian (\ref{typeIIL}) vanishes on-shell, $\,\cL_{\typeT}=0$,  up to the self-duality. \\


\section{Parametrization and  Gauge Fixing\label{SECpara}}
In this section,   taking   specific  parametrization   of the double-vielbein and  an $\ODD$ element,  we consider  a \textit{diagonal gauge fixing of the double  local Lorentz symmetries}.   We  discuss   the consequent   modification of the $\ODD$ transformation rule and  the flipping of the  chirality of the theory. We further show that   the gauge fixing  may map   the R-R potential  to an $\ODD$ spinor.  We refer readers  to  Appendix~\ref{SUBSECREDAPP} both  for the explicit  parametrization of the double-vielbein we are taking and  for  a self-contained review  of  Refs.\cite{Jeon:2011cn,Jeon:2011vx}  on the  reduction to Riemannian geometry  in  $D$ dimension.\\

\subsection{Parametrization of the $\ODD$ rotation} 
We parametrize         a generic  $\ODD$ group element,
\be
M_{A}{}^{B}=\left(\ba{cc}\mba^{\mu}{}_{\nu}&\mbb^{\mu\sigma}\\ 
\mbc_{\rho\nu}&\mbd_{\rho}{}^{\sigma}\ea\right)\,.
\label{Mpr}
\ee
The definition  of the  $\ODD$  group,
\be
\ba{ll}
M_{A}{}^{B}M_{C}{}^{D}\cJ_{BD}=\cJ_{AC}\,,~~~~&~~~~\cJ_{AB}=\left(\ba{cc}0&1\\1&0\ea\right)\,,
\ea
\ee
implies then
\be
\ba{llll}
\mba\mbb^{t}+\mbb\mba^{t}=0\,,~~~~&~~~~\mbc\mbd^{t}+\mbd\mbc^{t}=0\,,~~~~&~~~~\mba\mbd^{t}+\mbb\mbc^{t}=1\,.
\ea
\label{mbab}
\ee
From the vectorial $\ODD$ transformation rule of the double-vielbein,
\be
\ba{ll}
V_{Ap}~~\longrightarrow~~M_{A}{}^{B}V_{Bp}\,,~~~~&~~~~\brV_{A\brp}~~\longrightarrow~~M_{A}{}^{B}\brV_{B\brp}\,,
\ea
\label{vecunp}
\ee
the parametrization of the double-vielbein~(\ref{Vform1})  gives 
\be
\ba{ll}
e^{-1}~~\longrightarrow~~e^{-1}\left[\mba^{t}+(g-B)\mbb^{t}\right]\,,~~~~&~~~~
\bre^{-1}~~\longrightarrow~~\bre^{-1}\left[\mba^{t}-(g+B)\mbb^{t}\right]\,,
\ea
\label{ebreODD}
\ee
such that
\be
\ba{ll}
(e^{-1}\bre)_{p}{}^{\brp}~~\longrightarrow~~L_{p}{}^{q}(e^{-1}\bre)_{q}{}^{\brp}\,,~~~~&~~~~(\bre^{-1}e)_{\brp}{}^{p}~~\longrightarrow~~\brL_{\brp}{}^{\brq}(\bre^{-1}e)_{\brq}{}^{p}\,,
\ea
\label{eeLL}
\ee
where we set
\be
\ba{ll}
L=e^{-1}\left[\mba^{t}+(g-B)\mbb^{t}\right]\left[\mba^{t}-(g+B)\mbb^{t}\right]^{-1}e\,,
~~~~&~~~~
\brL=(\bre^{-1}e)L^{-1}(e^{-1}\bre)\,.
\ea
\label{LbrL}
\ee
From the considerations  that  $(e^{-1}\bre)_{p}{}^{\brp}$ and $(\bre^{-1}e)_{\brp}{}^{p}$ themselves  are local  Lorentz transformations and  that this property must be preserved under $\ODD$ T-duality rotation,  or alternatively  from direct verification using (\ref{mbab}),  a crucial property  of $L$ and $\brL$ follows:  they correspond to local Lorentz transformations, 
\be
\ba{ll}
L_{p}{}^{r}L_{q}{}^{s}\eta_{rs}=\eta_{pq}\,,~~~~&~~~~\brL_{\brp}{}^{\brr}\brL_{\brq}{}^{\brs}\breta_{\brr\brs}=\breta_{\brp\brq}\,.
\ea
\label{LLLorentz}
\ee
~\\

Even-dimensional irreducible gamma matrices are unique up to similarity transformations,  essentially   due to Schur's lemma. This implies,  for    (\ref{LLLorentz}), (\ref{gammabr}), (\ref{ebreeta}),  that   there must be similarity transformations, $S_{e}$ satisfying
\be
\ba{ll}
\brgamma^{\brp}(\bre^{-1}e)_{\brp}{}^{p}=S_{e}^{-1}(\gamma^{(D+1)}\gamma^{p})S_{e}\,,~~~~&~~~~\gamma^{(D+1)}\gamma^{p}(e^{-1}\bre)_{p}{}^{\brp}=S_{e}\brgamma^{\brp}S_{e}^{-1}\,,
\ea
\label{SgamS}
\ee
and   $S_{L}$, $S_{\brL}$ satisfying 
\be
\ba{ll}
\gamma^{q}L_{q}{}^{p}=S_{L}^{-1}\gamma^{p}S_{L}\,,~~~~&~~~~
\brgamma^{\brq}\brL_{\brq}{}^{\brp}=S_{\brL}^{-1}\brgamma^{\brp}S_{\brL}\,.
\ea
\label{gLS}
\ee
From  (\ref{gLS}), (\ref{g5}), we obtain 
\be
\ba{ll}
\gamma^{(D+1)}S_{L}=\det(L)\,S_{L}\gamma^{(D+1)}\,,
~~~~&~~~~\brgamma^{(D+1)}S_{\brL}=\det(\brL)\,S_{\brL}\brgamma^{(D+1)}\,,
\ea
\label{brS5}
\ee
where  from (\ref{LbrL}),
\be
\dis{
\det(L)=\det(\brL)^{-1}=\frac{\det\left[\mba+\mbb(g+B)\right]}{\det\left[\mba-\mbb(g-B)\right]}\,,}
\label{detL}
\ee
of which the  value must be either $+1$ or $-1$, since $L$ and $\brL$ are  local Lorentz transformations. 
Thus, if $\det(\brL)=+1$, $S_{\brL}$ commutes with $\brgamma^{(D+1)}$. Otherwise  \textit{i.e.~}$\det(\brL)=-1$,  they anti-commute~\cite{Jeon:2011vx}.\\

In fact, using (\ref{brS5}), one can show that $S_{L}$ and  $S_{\brL}$ are related by
\be
S_{\brL}=\left\{\ba{ll}
S_{e}^{-1}S_{L}^{-1}S_{e}\quad&\quad\mbox{for\,~}\det(\brL)=+1\\
S_{e}^{-1}\gamma^{(D+1)}S_{L}^{-1}S_{e}\quad&\quad\mbox{for\,~}\det(\brL)=-1\,.
\ea
\right.
\label{SLS}
\ee
~\\

For later use, we also parametrize an element of $\soDD$ Lie algebra: we set  a ${2D\times 2D}$ skew-symmetric matrix,
\be
h_{AB}=-h_{BA}=\left(\ba{cc}\alpha^{\mu\sigma}&-(\beta^{t})^{\mu}{}_{\rho}\\ \beta_{\nu}{}^{\sigma}&\gamma_{\nu\rho}\ea\right)
=\left(\ba{cc}-\alpha^{\sigma\mu}&  -\beta_{\rho}{}^{\mu}\\ \beta_{\nu}{}^{\sigma}&-\gamma_{\rho\nu}\ea\right)\,,
\label{hpara}
\ee
where $\alpha^{\mu\nu}$ and $\gamma_{\mu\nu}$ are arbitrary ${D\times D}$  skew-symmetric matrices, while $\beta_{\mu}{}^{\nu}$ is a generic ${D\times D}$ matrix. \\

From $M_{A}{}^{B}\approx \delta_{A}{}^{B}+h_{A}{}^{B}$ for (\ref{Mpr}), we obtain 
\be
\ba{ll}
L_{p}{}^{q}\approx \delta_{p}{}^{q}+\tau_{p}{}^{q}\,,~~~~&~~~~\tau= -2e^{-1}g\alpha e\,, \\

\brL_{\brp}{}^{\brq}\approx \delta_{\brp}{}^{\brq}+\brtau_{\brp}{}^{\brq}\,,~~~~&~~~~\brtau= +2\bre^{-1}g\alpha\bre\,,
\ea
\label{Ltau}
\ee
and hence, from (\ref{gLS}),
\be
\ba{ll}
S_{L}\approx 1-\quarter\tau_{pq}\gamma^{pq}\,,~~~~&~~~~
S_{\brL}\approx 1-\quarter\brtau_{\brp\brq}\brgamma^{\brp\brq}\,.
\ea
\ee
~\\

\subsection{Diagonal gauge fixing and modified $\ODD$ transformation rule}
Henceforth, we consider  a gauge fixing of the local Lorentz symmetries by setting
\be
e_{\mu}{}^{p}\equiv \bre_{\mu}{}^{\brp}\,.
\label{gaugefixing}
\ee
This gauge fixing breaks the double local Lorentz symmetry groups,   $\SpinD\times\oSpinD$,  to its diagonal subgroup, $\dSpinD$, which  acts   on the unbarred $\SpinD$ and the barred $\oSpinD$ indices simultaneously, reducing  SDFT to  ordinary supergravity~\cite{Jeon:2011sq}. \\

Moreover, from (\ref{ebreODD}),   the above diagonal gauge is  incompatible with  the  vectorial $\ODD$  transformation rule of the double-vielbein  (\ref{vecunp})~\cite{Jeon:2011cn,Jeon:2011vx}. In order to preserve the diagonal gauge, it is necessary to  modify  the $\ODD$ transformation rule: the $\ODD$ rotation  must accompany   a compensating  local Lorentz transformation. 
In   Table~\ref{TABmodODD}, we present  our   modified $\ODD$  T-duality transformation rule. It contains    a compensating  $\oPinD$  local Lorentz rotation given by  $\brL_{\brq}{}^{\brp}$  (\ref{LbrL}) and $S_{\brL}{}^{\bralpha}{}_{\brbeta}$ (\ref{gLS}),  which depend  on both the $\ODD$ element, $M$ (\ref{Mpr}), and the NS-NS  backgrounds,  $g_{\mu\nu}$, $B_{\mu\nu}$,
\be
\ba{ll}
\brL=\bre^{-1}\left[\mba^{t}-(g+B)\mbb^{t}\right]\left[\mba^{t}+(g-B)\mbb^{t}\right]^{-1}\bre\,,~~~~&~~~~
\brgamma^{\brq}\brL_{\brq}{}^{\brp}=S_{\brL}^{-1}\brgamma^{\brp}S_{\brL}\,.
\ea
\label{compL}
\ee
The modified $\ODD$ T-duality transformation rule implies, in particular, 
\be
\ba{lll}
\quad\cF~~&~\longrightarrow~&~\cF S_{\brL}^{-1}\,.
\ea
\ee

\begin{table}[H]
\begin{center}
\be
\boxed{
\ba{clc}
\quad d~~&~\longrightarrow~&~d~~~\\
\quad V_{A}{}^{p}~~&~\longrightarrow~&~M_{A}{}^{B\,}V_{B}{}^{p}~~~ \\
\quad \brV_{A}{}^{\brp}~~&~\longrightarrow~&~M_{A}{}^{B\,}\brV_{B}{}^{\brq\,}\brL_{\brq}{}^{\brp}~~~ \\
\quad\cC~~&~\longrightarrow~&~\cC S_{\brL}^{-1}~~~\\
\quad\rho~~&~\longrightarrow~&~\rho~~~\\
\quad\rho^{\prime}~~&~\longrightarrow~&~S_{\brL}\rho^{\prime}~~~\\
\quad\psi_{\brp}~~&~\longrightarrow~&~(\brL^{-1})_{\brp}{}^{\brq\,}\psi_{\brq}~~~\\
\quad\psi^{\prime}_{p}~~&~\longrightarrow~&~S_{\brL}\psi_{p}^{\prime}~~~
\ea}
\label{avecp}
\ee
\caption{Modified $\ODD$  T-duality transformation rule, after  the diagonal gauge fixing  (\ref{gaugefixing}).   It contains  an induced    $\oPinD$ local Lorentz rotation~(\ref{compL})  (\textit{cf.~}\cite{Hassan:1999bv,Hassan:1999mm,Hassan:2000kr} and  also  \cite{Sfetsos:2010uq,Lozano:2011kb,Sfetsos:2011jw}).}
\label{TABmodODD}
\end{center}
\end{table}

Clearly, from (\ref{brS5}), if and only if  $\det(\brL)=-1$,  the modified $\ODD$ rotation  flips the chirality    of the  R-R potential (\ref{RRDS}) as well as those of the  primed fermions, $\rhop$, $\psi^{\prime}_{p}$ (\ref{chrialityfermion}), since
\be
\ba{lll}
\gamma^{(D+1)}\cC\brgamma^{(D+1)}=\pm\cC~~&~\longrightarrow~&~~
\gamma^{(D+1)}(\cC S_{\brL}^{-1})\brgamma^{(D+1)}=\pm\det(\brL)(\cC S_{\brL}^{-1})\,,\\
\brgamma^{(D+1)}\psi^{\prime}_{p}=\pm\psi^{\prime}_{p}~~&~\longrightarrow~&~~
\brgamma^{(D+1)}(S_{\brL}\psi^{\prime}_{p})=\pm\det(\brL)(S_{\brL}\psi^{\prime}_{p})\,,\\
\brgamma^{(D+1)}\rhop=\mp\rhop~~&~\longrightarrow~&~~
\brgamma^{(D+1)}(S_{\brL}\rhop)=\mp\det(\brL)(S_{\brL}\rhop)\,.
\ea
\ee

Thus, the mechanism above  naturally realizes   the  exchange of   type IIA and IIB supergravities under $\ODD$ T-duality within DFT setup, as anticipated in \cite{Jeon:2011vx}.\footnote{While  our  analysis of the chirality change    is technically based on our earlier  work on DFT fermions~\cite{Jeon:2011vx},  novel contributions of the present paper include  its generalization to R-R sector and  the realization that the diagonal gauge fixing  inevitably calls for   the modification of the $\ODD$  transformation rule.}  
For example, on a flat background  ($g=\eta$, $B=0$), we may set both $\mba$ and $\mbb g$ to be diagonal with  the eigenvalues,  either zero or one, in an exclusive manner such that  $\mba+\mbb g=1$. This choice corresponds to the usual discrete T-duality along toroidal directions.  In this case, we get  $\det(\brL)=(-1)^{\sharp_{\mba}}$ where $\sharp_{\mba}$ counts the number of zero  eigenvalues in the matrix, $\mba$,  and hence the number of toroidal directions on which T-duality is performed.  Thus, our formula is consistent with the well-known knowledge   that performing odd number of T-duality on flat backgrounds  exchanges type IIA and IIB superstrings.\\

It is also worth while to note that, since the compensating local Lorentz rotation~(\ref{compL}) explicitly depends on the parametrization of the double-vielbein in terms of   $g_{\mu\nu}$ and $B_{\mu\nu}$,  it appears   impossible to impose    the modified $\ODD$ transformation rule, Table~\ref{TABmodODD}, from the beginning  in the parametrization-independent covariant formalism.  \\

For later use, we write down  the modified infinitesimal  $\soDD$ transformation rule, 
\be
\ba{ll}
\hdelta_{h} d=0\,,~~~~\quad&~~~~\hdelta_{h}\cC=\quarter\brtau_{\brp\brq}\cC\brgamma^{\brp\brq}\,,\\
\hdelta_{h} V_{A}{}^{p}=h_{A}{}^{B}V_{B}{}^{p}\,,~~~~\quad&~~~~
\hdelta_{h}\brV_{A}{}^{\brp}=h_{A}{}^{B\,}\brV_{B}{}^{\brp}+\brV_{A}{}^{\brq}\brtau_{\brq}{}^{\brp}\,, \\
\hdelta_{h}\rho=0\,,~~~~\quad&~~~~
\hdelta_{h}\rho^{\prime}=-\quarter\brtau_{\brp\brq}\brgamma^{\brp\brq}\rho^{\prime}\,,\\
\hdelta_{h}\psi_{\brp}=-\brtau_{\brp}{}^{\brq}\psi_{\brq}\,,~~~~\quad&~~~~
\hdelta_{h}\psi^{\prime}_{p}=-\quarter\brtau_{\brp\brq}\brgamma^{\brp\brq}\psi_{p}^{\prime}\,,
\ea
\label{hdeltah}
\ee
and  
\be
\ba{l}
\hdelta_{h} e_{\mu}{}^{p}= (\beta_{\mu}{}^{\nu}-B_{\mu\rho}\alpha^{\rho\nu}+g_{\mu\rho}\alpha^{\rho\nu})e_{\nu}{}^{p}\,,\\
\hdelta_{h} \bre_{\mu}{}^{\brp}= (\beta_{\mu}{}^{\nu}-B_{\mu\rho}\alpha^{\rho\nu}+g_{\mu\rho}\alpha^{\rho\nu})\bre_{\nu}{}^{\brp}\,,\\
\hdelta_{h} B_{\mu\nu}= \gamma_{\mu\nu}-2\beta_{[\mu}{}^{\rho}B_{\nu]\rho}-g_{\mu\rho}\alpha^{\rho\sigma} g_{\sigma\nu}-B_{\mu\rho}\alpha^{\rho\sigma} B_{\sigma\nu}\,,\\
\hdelta_{h}\phi=\half(\beta_{\mu}{}^{\mu}-B_{\mu\rho}\alpha^{\rho\mu})\,,
\ea
\label{hdeltaeeB}
\ee
where we put, ${\hdelta_{h}=\delta_{h}-y^{A}h_{A}{}^{B}\partial_{B}}$  for our short hand notation: $\delta_{h}$ is the actual transformation and the derivative,  ${y^{A}h_{A}{}^{B}\partial_{B}}$, is for   the  transformation of the coordinates, $\delta_{h}y^{A}=y^{B}h_{B}{}^{A}$.\\
~\\

\subsection{Mapping the R-R potential to  an $\ODD$ spinor}
In this subsection,  we show that   after  the diagonal gauge fixing~(\ref{gaugefixing}), the modified $\ODD$ transformation rule  of   the    R-R potential in Table~\ref{TABmodODD} actually  implies that  the $\SpinD\times\oSpinD$  bi-fundamental R-R potential can be mapped  to  an $\ODD$ spinor.  \\

With the gauge fixing, $e_{\mu}{}^{p}\equiv\bre_{\mu}{}^{\brp}$~(\ref{gaugefixing}), from (\ref{LbrL}), (\ref{Ltau}), we have
\be
\ba{ll}
\brL\equiv L^{-1}\,,~~~~&~~~~\tau_{p}{}^{q}\equiv -\brtau_{\brp}{}^{\brq}\,.
\ea
\ee
Further, we may set
\be
\ba{lll}
\eta_{pq}\equiv -\breta_{\brp\brq}\,,~~~~&~~~~\brgamma^{\brp}\equiv\gamma^{(D+1)}\gamma^{p}\,,~~~~&~~~~C_{+\alpha\beta}\equiv\brC_{+\bralpha\brbeta}\,,
\ea
\ee
such that, from  (\ref{g5}), (\ref{omega2}),
\be
\ba{llll}
\brgamma^{(D+1)}\equiv-\gamma^{(D+1)}\,,~~~&~~~\tau_{pq}\gamma^{pq}\equiv-\brtau_{\brp\brq}\brgamma^{\brp\brq}\,,~~~&~~~\omega_{\mu p}{}^{q}\equiv\bromega_{\mu\brp}{}^{\brq}\,,~~~&~~~
\omega_{\mu pq}\gamma^{pq}\equiv\bromega_{\mu\brp\brq}\brgamma^{\brp\brq}\,.
\ea
\ee
~\\

As the diagonal gauge fixing tends to eliminate     the distinction of  the two local Lorentz groups, we may expand the R-R potential by one sort of  gamma matrices,  as in the democratic formulation of the R-R sector~\cite{Townsend:1995gp,Green:1996bh,Bergshoeff:2001pv},
\be
\cC\equiv \left(\half\right)^{\frac{D+2}{4}}\textstyle{\sum^{\prime}_{p}}\,\textstyle{\frac{1}{p !}}\,\cC_{a_{1}a_{2} \cdots a_p}\gamma^{a_{1}a_{2} \cdots a_p}\,.
\label{biRR}
\ee
where  $\sum^{\prime}_{p}$ denotes the odd $p$ sum for type IIA and even $p$ sum for type IIB.\\

Consequently the field   strength reads
\be
\cF=\cDo_{+}\cC\equiv \left(\half\right)^{\frac{D}{4}}\textstyle{\sum^{\prime}_{p}}\,\textstyle{\frac{1}{(p+1)!}}\,\cF_{a_1a_{2}\cdots a_{p+1}}\gamma^{a_1a_{2} \cdots a_{p+1}}\,,
\ee
where,  up to the section choice~(\ref{sectionchoice}),  with $D_{\mu}\equiv\trd_{\mu}+\omega_{\mu}$ from (\ref{DD}),
\be
\cF_{a_1a_{2} \cdots a_{p}}\simeq p\left(D_{[a_1}\cC_{a_2 \cdots a_{p}]}-\partial_{[a_1}\phi \,\cC_{a_2 \cdots a_{p}]}\right)+\textstyle{\frac{p !}{3!(p-3)!}}\,H_{[a_1 a_2 a_3}\cC_{a_4 \cdots a_{p}]}\,.
\ee
Similarly we also obtain
\be
\cDo_{-} \cC\simeq\textstyle{\sum^{\prime}_{p}}\,\textstyle{\frac{1}{(p-1)!}}\left(D^{b}\cC_{ba_1 \cdots a_{p-1}}-\partial^{b}\phi\, \cC_{ba_1 \cdots a_{p-1}}-\frac{1}{3!}H^{bcd}\cC_{bcda_{1}\cdots a_{p-1}}\right)\gamma^{a_1 \cdots a_{p-1}}\,.
\ee
That is to say, the pair of   nilpotent differential operators, $\cDo_{+}$ and $\cDo_{-}$,  reduce    to an exterior derivative and its  dual derivative respectively,
\be
\ba{lll}
\cDo_{+}\quad&~\Longrightarrow~&\quad~~\rmd +(H-{\rmd}\phi)\wedge\quad\quad~~,\\
\cDo_{-}\quad&~\Longrightarrow~&\quad\ast\left[\,\rmd +(H-{\rmd}\phi)\wedge~\right]\ast~\,.
\ea
\ee
~\\

The R-R sector Lagrangian~(\ref{RRL})   assumes   the  standard form now,
\be
\cL_{\RR}=-\half e^{-2d}\Tr(\cF \bar{\cF})\equiv-\half e^{-2d}\textstyle{\sum^{\prime}_{p}}\,\textstyle{\frac{1}{(p+1)!}}\cF_{a_1a_{2} \cdots a_{p+1}}\cF^{a_{1}a_{2}\cdots a_{p+1}}\,.
\ee
And  the infinitesimal $\soDD$ transformation of the R-R potential~(\ref{hdeltah}),
\be
\hdelta_{h}\cC =\quarter\bar{\tau}_{\bra\brb}\cC\brgamma^{\bra\brb}\equiv-\quarter\tau_{ab}\cC\gamma^{ab}\,,
\ee 
gives the   transformation of each $p$-form potential, 
\be
\hdelta_{h}\cC_{a_1 \cdots a_p}=-\quarter\tau_{[a_1 a_2}\cC_{a_3 \cdots a_p]}+\half p\tau_{[a_1}{}^{b}\cC_{|b |a_2 \cdots a_p]}+\quarter\tau^{bc}\cC_{bc a_1 \cdots a_p}\,.
\label{hdeltacC}
\ee
~\\

We continue to  consider   a formal sum of the $p$-forms,
\be
\ba{ll}
\hat{\cC}:=\textstyle{\sum^{\prime}_{p}}\,\textstyle{\frac{1}{p!}}\,\cC_{\mu_1  \cdots \mu_p}{\rmd x}^{\mu_1}\wedge \cdots \wedge {\rm d}x^{\mu_p}\,,~~~~&~~~~
\cC_{\mu_1  \cdots \mu_p}=e_{\mu_{1}}{}^{a_{1}}\cdots e_{\mu_{p}}{}^{a_{p}}\cC_{a_1  \cdots a_p}\,,
\ea
\ee
and perform  a field redefinition of the formal sum,
\be
\hat{\cA}:=e^{-\phi}e^{B}\wedge \hat{\cC}\,.
\label{map}
\ee
From its expansion,
\be
\hat{\cA}=\textstyle{\sum^{\prime}_{p}}\,\textstyle{\frac{1}{p!}}\,\cA_{\mu_1\cdots \mu_p}{\rmd x}^{\mu_1}\wedge \cdots \wedge  {\rm d}x^{\mu_p}\,,
\ee
we finally  define an $\ODD$ spinor,
\be
|\cA\rangle := \textstyle{\sum^{\prime}_{p}}\,\textstyle{\frac{1}{p!}}\,\cA_{\mu_1\mu_{2} \cdots \mu_p} \left(\textstyle{\frac{1}{\sqrt{2}}}\Gamma^{\mu_1}\right)\left(\textstyle{\frac{1}{\sqrt{2}}}\Gamma^{\mu_2}\right)\cdots \left(\textstyle{\frac{1}{\sqrt{2}}}\Gamma^{\mu_p}\right)|0\rangle\,.
\label{ODDRR}
\ee
Here $\frac{1}{\sqrt{2}}\Gamma^{\mu}$'s are the normalized creation gamma matrices~\cite{Fukuma:1999jt,Hohm:2011zr,Hohm:2011dv}. Together with the normalized  annihilation gamma matrices, $\frac{1}{\sqrt{2}}\Gamma_{\mu}$, they form   $\ODD$ Clifford algebra,
\be
\ba{ll}
\Gamma^{A}=\left(\Gamma_{\mu}, \Gamma^{\nu}\right)\,,~~~~&~~~~
\Gamma^{A}\Gamma^{B}+\Gamma^{B}\Gamma^{A}=2\cJ^{AB}\,.
\ea
\ee
The annihilation gamma matrices annihilate the vacuum,  $\Gamma_{\mu}|0\rangle=0$.  It is worth while to note that,    unlike  the $D$-dimensional gamma matrices satisfying  $\gamma^{\mu}=\gamma^{p}(e^{-1})_{p}{}^{\mu}$,  the $\ODD$ gamma matrices, $\Gamma^{A}=(\Gamma_{\mu}, \Gamma^{\nu})$,  are all constant.\\

From    (\ref{hdeltaeeB}), (\ref{hdeltacC}), the infinitesimal  $\soDD$ transformation of the $p$-form, $\cA _{\mu_1 \cdots \mu_p}$, follows  
\be\ba{ll}
\hat{\delta}\cA _{\mu_1 \cdots \mu_p}=&-\half \beta_{\lambda}{}^{\lambda}\cA_{\mu_1 \cdots \mu_p} +\frac{p(p-1)}{2}\gamma_{[\mu_1 \mu_2}\cA_{\mu_3 \cdots \mu_p]}+p\beta_{[\mu_1}{}^{\nu}\cA_{|\nu| \mu_2 \cdots \mu_p]}-\half \alpha^{\nu\lambda}\cA_{\nu\lambda \mu_1 \cdots \mu_p}\,.
\ea\label{Abasisodd}
\ee
It is straightforward to check that, this result is equivalent to the  $\soDD$ transformation of the $\ODD$   spinor~(\ref{ODDRR}),
 \be
\hdelta_{h}|\cA\rangle = \quarter h_{AB}\Gamma^{AB}|\cA\rangle\,,
\ee
establishing   the desired result,
\be
\ba{lll}
\hdelta_{h}\cC =\quarter\bar{\tau}_{\bra\brb}\cC\brgamma^{\bra\brb}&\quad\Longleftrightarrow\quad&
\hdelta_{h}|\cA\rangle = \quarter h_{AB}\Gamma^{AB}|\cA\rangle\,.
\ea
\label{equivCA}
\ee
This completes our proof.\\

\section{Conclusion\label{SECCON}}
In this work, we have  incorporated the R-R sector into double field theory in a manifestly covariant manner, with respect to $\ODD$ T-duality, double-gauge symmetry and the pair of local Lorentz symmetries, $\SpinD\times\oSpinD$.  We  put the R-R sector in the  bi-fundamental spinorial   representation of the double   Lorentz groups, and constructed a pair of nilpotent differential operators, (\ref{DpmS}): one for the field strength (\ref{cFdef}) and other for the equation of motion (\ref{EOMRR}). \\

We have spelled  out the bosonic part of the type II supersymmetric double field theory Lagrangian (\ref{typeIIL}). We  presented  the equations of motion (\ref{LNSd}),  (\ref{EinsteinDFT}), (\ref{EOMRR2}),  and analyzed   the self-duality of the field strength (\ref{SD}) .\\

\textit{A priori},  in the  parametrization independent covariant formalism (section~\ref{SECRR} and Appendix~\ref{SECconv}),    the R-R sector and  all the fermions are  $\ODD$ singlet.  Yet,  after gauge fixing the two vielbeins  equal to each other breaks the 
double local Lorentz groups to the diagonal subgroup, 
\be
\SpinD\times\oSpinD\,\rightarrow\,\dSpinD\,.
\ee
Further, it modifies the $\ODD$ transformation  rule to call for   a compensating $\oPinD$ rotation, (\ref{compL}), which flips  the chirality of the theory,     if and only if $\det(\brL)=-1$ (\ref{detL}),   resulting in the exchange of type IIA and IIB supergravities.  The modified $\soDD$ transformation rule of  the R-R potential  can be  mapped to  an $\soDD$  rotation of a corresponding  $\ODD$ spinor~(\ref{ODDRR}), (\ref{equivCA}). \\

We emphasize that,  the equivalence~(\ref{equivCA}) between     the  double Lorentz bi-fundamental   treatment of the R-R sector, $\cC^{\alpha}{}_{\bralpha}$,  and the $\ODD$ spinorial  treatment of it, $|\cA\rangle$, holds only after taking the  diagonal gauge fixing~(\ref{gaugefixing}).  The parametrization independent covariant formalism (section  \ref{SECRR} and Appendix \ref{SECconv}) appears to prefer  the former approach. As mentioned ahead, while the former may couple to fermions naturally,   the latter appears rather awkward to do so.  The ${\cN=2}$ supersymmetrization of the Lagrangian~(\ref{typeIIL}) remains as a future work~\cite{preparation}.\\

\section*{Acknowledgements} We wish to thank  David Geissbuhler and  Diego  Marqu\'es   for interesting  discussions. We are indebted to   Yoonji Suh for proofreading of the manuscript.     The work was supported by the National Research Foundation of Korea\,(NRF) grants  funded by the Korea government\,(MEST) with the Grant No.  2005-0049409 (CQUeST)  and No.  2010-0002980.  \newpage


\appendix
\section{Review: Stringy Differential Geometry for  covariant  SDFT\label{SECconv}}

\subsection{Fundamental field variables  and  conventions}
In   Table~\ref{TABfields}, we spell out all the fundamental   field variables which  constitute     type II  (or {${\cN=2}$ })    SDFT.  
In Table \ref{TABindices}, we  summarize our conventions for  indices and   metrics  which are being used  for each  representation of the symmetries    in Table \ref{TABsymmetry}.

\begin{table}[H]
\begin{center}
\begin{itemize}
\item Bosons 
\begin{itemize}
\item NS-NS sector $~\left\{
\ba{lll}
\mbox{DFT-dilaton:}~~~&~~~\,d\\
\mbox{Double-vielbeins:}~~~&~~~V_{Ap}\,,~~\quad\brV_{A\brp}
\ea
\right.$
\item  R-R potential:   $\,~~\quad\quad \quad\quad\quad\quad\quad\quad\quad\quad \cC^{\alpha}{}_{\bralpha}$
\end{itemize}
\item Fermions
\begin{itemize}
\item DFT-dilatinos:  $\quad\quad\quad\quad \quad\quad\quad\quad \quad\quad\quad\rho^{\alpha}\,$,  $\,~~~\quad\rho^{\prime\bralpha}$
\item Gravitinos:~~ $~~~\quad\quad\quad\quad \quad\quad\quad\quad \quad\quad\quad\psi^{\alpha}_{\brp}\,$, $~~\quad\psi^{\prime\bralpha}_{p}$ 
\end{itemize}
\end{itemize}
\caption{Fundamental  field variables in  type II SDFT}
\label{TABfields}
\end{center}
\end{table}

\begin{table}[H]
\begin{center}
\begin{tabular}{c|c|c}
~~~Index~~~&~~Representation~~&~~~Metric~~~\\
\hline
$A,B,\cdots$~&~$\ODD$ \& double-gauge vector~&$\cJ_{AB}$ in Eq.(\ref{ODDeta})\\
$p,q,\cdots$~&~$\SpinD\,$  vector~&$\eta_{pq}=\mbox{diag}(-++\cdots+)$ \\
$\alpha,\beta,\cdots$~&~$\SpinD\,$  spinor~&
$\Cp_{\alpha\beta}$  in Eq.(\ref{CcM})\\
$\brp,\brq,\cdots$~&~$\oSpinD\,$  vector~&$\breta_{\brp\brq}=\mbox{diag}(+--\cdots-)$ \\
$\bralpha,\brbeta,\cdots$~&~$\oSpinD\,$  spinor~&
$\brCp_{\bralpha\brbeta}$  in Eq.(\ref{CcM})\\
\end{tabular}
\caption{Indices used for each symmetry representation and the relevant    metrics that raise or lower the positions of them. \textit{A priori},  $\ODD$ rotates   only  the double-gauge vector  indices (capital Roman). All the metrics have been  chosen to be symmetric in this paper, \textit{cf.}~\cite{Jeon:2011vx,Jeon:2011sq}.}
\label{TABindices}
\end{center}
\end{table}
Throughout the  paper we focus on even $D$-dimensional Minkowskian  spacetime  which  admits  Majorana-Weyl spinors, and hence $D\equiv  2$ mod $8$,  or simply ${D=10}$ for the critical superstring theory.\\ 

For the two Minkowskian  metrics, $\eta_{pq}$ and $\breta_{\brp\brq}$,   we introduce  separately the corresponding real gamma matrices: $(\gamma^{p})^{\alpha}{}_{\beta}$ and $(\brgamma^{\brp})^{\bralpha}{}_{\brbeta}$ satisfying 
\be
\ba{ll}
\gamma^{p}=(\gamma^{p})^{\ast}\,,~~~~&~~~~
\gamma^{p}\gamma^{q}+\gamma^{q}\gamma^{p}=2\eta^{pq}\,,\\
\brgamma^{\brp}=(\brgamma^{\brp})^{\ast}\,,~~~~&~~~~
\brgamma^{\brp}\brgamma^{\brq}+\brgamma^{\brq}\brgamma^{\brp}=2\breta^{\brp\brq}\,.
\ea
\label{gammabr}
\ee  
Their  charge conjugation matrices,  $C_{\pm}{}_{\alpha\beta}$ and $\brC_{\pm}{}_{\bralpha\brbeta}$, meet
\be
\ba{l}
(\Cp\gamma^{p_{1}p_{2}\cdots p_{n}})_{\alpha\beta}=(-1)^{n(n-1)/2}(\Cp\gamma^{p_{1}p_{2}\cdots p_{n}})_{\beta\alpha}\,,\\
(\brCp\brgamma^{\brp_{1}\brp_{2}\cdots\brp_{n}})_{\bralpha\brbeta}=(-1)^{n(n-1)/2}(\brCp\brgamma^{\brp_{1}\brp_{2}\cdots\brp_{n}})_{\brbeta\bralpha}\,,
\ea
\label{CcM}
\ee
and define the charge-conjugated  spinors,\footnote{In this work, we have changed  our   convention   in the definition of    the charge-conjugation    from the previous works~\cite{Jeon:2011vx,Jeon:2011sq},  such that   we employ  not 
the anti-symmetric charge conjugation matrices, $\Cm_{\alpha\beta}=-\Cm_{\beta\alpha}$ and  $\brCm_{\bralpha\brbeta}=-\brCm_{\brbeta\bralpha}$ but the symmetric charge conjugation matrices, $\Cp_{\alpha\beta}=\Cp_{\beta\alpha}$ and  $\brCp_{\bralpha\brbeta}=\brCp_{\brbeta\bralpha}$  (\ref{cgspinor}), in order to reduce the number of minus signs appearing in the actual  computations.   They are related by 
\[
\ba{ll}
\Cm=\Cp\gamma^{(D+1)}\,,~~~~&~~~~\brCm=\brCp\brgamma^{(D+1)}\,.
\ea
\]
The  anti-symmetric charge conjugation matrices satisfy
\[
\ba{l}
(\Cm\gamma^{p_{1}p_{2}\cdots p_{n}})_{\alpha\beta}=-(-1)^{n(n+1)/2}(\Cm\gamma^{p_{1}p_{2}\cdots p_{n}})_{\beta\alpha}\,,\\
(\brCm\brgamma^{\brp_{1}\brp_{2}\cdots\brp_{n}})_{\bralpha\brbeta}=-(-1)^{n(n+1)/2}(\brCm\brgamma^{\brp_{1}\brp_{2}\cdots\brp_{n}})_{\brbeta\bralpha}\,.
\ea
\]
}
\be
\ba{ll}
\brpsi_{\brp\alpha}=\psi_{\brp}^{\,\beta}\Cp_{\beta\alpha}\,,~~~~&~~~~
\brrho_{\alpha}=\rho^{\beta} \Cp_{\beta\alpha}\,,\\
\brpsi^{\prime}_{p\bralpha}=\psi_{p}^{\prime\brbeta}\brCp_{\brbeta\bralpha}\,,~~~~&~~~~
\brrho^{\prime}_{\bralpha}=\rho^{\prime\brbeta}\brCp_{\brbeta\bralpha}\,.
\ea
\label{cgspinor}
\ee
\\
We also set, in order to specify  the chirality of the Weyl spinors,
\be
\ba{ll}
\gamma^{(D+1)}:=\gamma^{012\cdots D-1}\,,~~~~&~~~~\brgamma^{(D+1)}:=\brgamma^{012\cdots D-1}\,,
\ea
\label{g5}
\ee
which satisfy
\be
\ba{ll}
\gamma^{p}\gamma^{(D+1)}+\gamma^{(D+1)}\gamma^{p}=0\,,~~~&~~~~\left(\gamma^{(D+1)}\right)^{2}=1\,,\\\brgamma^{\brp}\brgamma^{(D+1)}+\brgamma^{(D+1)}\brgamma^{\brp}=0\,,~~~&~~~~\left(\brgamma^{(D+1)}\right)^{2}=1\,.
\ea
\ee
~\\

Specifically, the type II SDFT field variables in Table~\ref{TABfields} satisfy the following properties.
\begin{itemize}
\item The DFT-dilaton gives rise to a scalar density  with  weight one,
\be
e^{-2d}\,.
\ee 
\item The double-vielbeins  satisfy the  defining properties~\cite{Jeon:2011cn}:
\be
\ba{llll}
V_{Ap}V^{A}{}_{q}=\eta_{pq}\,,~~~&~~~~
\brV_{A\brp}\brV^{A}{}_{\brq}=\breta_{\brp\brq}\,,\\
V_{Ap}\brV^{A}{}_{\brq}=0\,,~~~&~~~~V_{Ap}V_{B}{}^{p}+\brV_{A\brp}\brV_{B}{}^{\brp}=\cJ_{AB}\,.
\ea
\label{defV}
\ee
\item The R-R potential, $\cC^{\alpha}{}_{\bralpha}$,  is in the bi-fundamental  spinorial   representation of the local Lorentz group,  $\SpinD\times\oSpinD$, and satisfies a `chirality' condition,
\be
\gamma^{(D+1)}\cC\brgamma^{(D+1)}=\pm\,\cC\,.
\label{RRD}
\ee
The upper sign is for type IIA and the lower  sign is for   type IIB.
\item  The unprimed fermions (R-NS), $\psi_{\brp}^{\,\alpha}$, $\rho^{\alpha}$, are  set to be  Majorana-Weyl spinors of the  fixed chiralities, 
\be
\ba{ll}
\gamma^{(D+1)}\psi_{\brp}=+\psi_{\brp}\,,~~~~&~~~~\gamma^{(D+1)}\rho=-\rho\,,
\ea
\label{chiralityunp}
\ee
while the primed   fermions (NS-R), $\psi_{p}^{\prime\bralpha}$, $\rho^{\prime\bralpha}$, are  Majorana-Weyl spinors of a definite   yet unfixed chirality,
\be
\ba{ll}
\brgamma^{(D+1)}\psi^{\prime}_{p}=\pm\psi^{\prime}_{p}\,,~~~~&~~~~\brgamma^{(D+1)}\rhop=\mp\rhop\,.
\ea
\label{chrialityfermion}
\ee
Again, the upper sign is for  type IIA and the lower  sign is for type IIB. This somewhat unconventional  IIA/IIB identification,  compared with (\ref{chiralityunp}),  is  due to the opposite signatures of  the $D$-dimensional metrics, $\eta$ and $\breta$,  we have chosen   in Table~\ref{TABindices}.
\end{itemize}

The double-vielbeins then generate a pair of rank-two projections~\cite{Jeon:2010rw},
\be
\ba{ll}
P_{AB}:=V_{A}{}^{p}V_{Bp}\,,~~~~&~~~~P_{A}{}^{B}P_{B}{}^{C}=P_{A}{}^{C}\,,\\
\brP_{AB}:=\brV_{A}{}^{\brp}\brV_{B\brp}\,,~~~~&~~~~\brP_{A}{}^{B}\brP_{B}{}^{C}=\brP_{A}{}^{C}\,,
\ea
\ee
and further a pair of rank-six projections~\cite{Jeon:2011cn}, 
\be
\ba{ll}
\cP_{CAB}{}^{DEF}:=P_{C}{}^{D}P_{[A}{}^{[E}P_{B]}{}^{F]}+\textstyle{\frac{2}{D-1}}P_{C[A}P_{B]}{}^{[E}P^{F]D}\,,~~&~~{\cP_{CAB}{}^{DEF}\cP_{DEF}{}^{GHI}=\cP_{CAB}{}^{GHI}\,,}\\
\bcP_{CAB}{}^{DEF}:=\brP_{C}{}^{D}\brP_{[A}{}^{[E}\brP_{B]}{}^{F]}+\textstyle{\frac{2}{D-1}}\brP_{C[A}\brP_{B]}{}^{[E}\brP^{F]D}\,,~~&~~{\bcP_{CAB}{}^{DEF}\bcP_{DEF}{}^{GHI}=\bcP_{CAB}{}^{GHI}\,.}
\ea
\label{P6}
\ee
The rank-two projections   are symmetric, orthogonal and complementary to each other,  
\be
\ba{llll}
P_{AB}=P_{BA}\,,~~&~~\brP_{AB}=\brP_{BA}\,,~~&~~P_{A}{}^{B}\brP_{B}{}^{C}=0\,,
~~&~~P_{A}{}^{B}+\brP_{A}{}^{B}=\delta_{A}{}^{B}\,,
\ea
\label{symP2}
\ee
satisfying 
\be
\ba{llll}
P_{A}{}^{B}V_{Bp}=V_{Ap}\,,~~&~~\brP_{A}{}^{B}\brV_{B\brp}=\brV_{A\brp}\,,~~&~~\brP_{A}{}^{B}V_{Bp}=0\,,~~&~~P_{A}{}^{B}\brV_{B\brp}=0\,.
\ea
\ee
The rank-six projections   are  symmetric and traceless,
\be
\ba{ll}
{\cP_{CABDEF}=\cP_{DEFCAB}=\cP_{C[AB]D[EF]}\,,}~~&~~{\bcP_{CABDEF}=\bcP_{DEFCAB}=\bcP_{C[AB]D[EF]}\,,} \\
{\cP^{A}{}_{ABDEF}=0\,,~~~~\,P^{AB}\cP_{ABCDEF}=0\,,}~~&~~
{\bcP^{A}{}_{ABDEF}=0\,,~~~~\,\brP^{AB}\bcP_{ABCDEF}=0\,.}
\ea
\label{symP6}
\ee
~\\

 For simplicity, we let 
\be
\ba{ll}
\psi_{A}:=\brV_{A}{}^{\brp}\psi_{\brp}\,,~~~~&~~~~\psi^{\prime}_{A}:=V_{A}{}^{p}\psi^{\prime}_{p}\,,\\
\gamma^{A}:=V^{A}{}_{p}\gamma^{p}\,,~~~~&~~~~\brgamma^{A}:=\brV^{A}{}_{\brp}\brgamma^{\brp}\,,
\ea
\label{psigamma}
\ee
such  that 
\be
\ba{ll}
{}\,~\brV^{A}{}_{\brp}\psi_{A}=\psi_{\brp}\,,~~~~&~~~~{}\,~V^{A}{}_{p}\psi^{\prime}_{A}=\psi^{\prime}_{p}\,,\\
{}\,~V^{A}{}_{p}\psi_{A}=0\,,~~~~&~~~~{}\,~\brV^{A}{}_{\brp}\psi^{\prime}_{A}=0\,,\\
\left\{\gamma^{A},\gamma^{B}\right\}=2P^{AB}\,,~~~~&~~~~\left\{\brgamma^{A},\brgamma^{B}\right\}=2\brP^{AB}\,.
\ea
\ee 
~\\

\subsection{Semi-covariant derivatives}
For each of  the  DFT gauge symmetry  in  Table~\ref{TABsymmetry},  we assign a corresponding  connection,  
\begin{itemize}
\item $~\Gamma_{A}$   for the double-gauge symmetry, 
\item $~\Phi_{A}$ for the  `unbarred' local    Lorentz  symmetry, $\SpinD$,
\item  $~\bar{\Phi}_{A}$ for the `barred'  local Lorentz symmetry, $\oSpinD$.
\end{itemize}
Combining all of them, we employ   the \textit{{master semi-covariant derivative}}~\cite{Jeon:2011vx},
\be
\cD_{A}=\partial_{A}+\Gamma_{A}+\Phi_{A}+\brPhi_{A}\,.
\label{cDA}
\ee
It is also useful to set 
\be
\ba{ll}
\na_{A}=\partial_{A}+\Gamma_{A}\,,~~~~&~~~~D_{A}=\partial_{A}+\Phi_{A}+\brPhi_{A}\,,
\ea
\label{nadd}
\ee
of which the former is the  semi-covariant derivative   for the double-gauge symmetry~\cite{Jeon:2010rw,Jeon:2011cn},
\be
\na_{C}\Tw_{A_{1}A_{2}\cdots A_{n}}
:=\partial_{C}\Tw_{A_{1}A_{2}\cdots A_{n}}-\omega_{{\scriptscriptstyle{T\,}}}\Gamma^{B}{}_{BC}\Tw_{A_{1}A_{2}\cdots A_{n}}+
\sum_{i=1}^{n}\,\Gamma_{CA_{i}}{}^{B}\Tw_{A_{1}\cdots A_{i-1}BA_{i+1}\cdots A_{n}}\,.
\label{semi-covD}
\ee
And the latter is  the covariant derivative for  the pair of  local Lorentz symmetries, yet being  semi-covariant under  the double-gauge symmetry~\cite{Jeon:2011vx}. \\

Firstly, the master semi-covariant derivative  is compatible with  all the constant metrics, 
 \be
 \ba{l}
\cD_{A}\cJ_{BC}=\na_{A}\cJ_{BC}=\Gamma_{AB}{}^{D}\cJ_{DC}+\Gamma_{AC}{}^{D}\cJ_{BD}=0\,,\\
\cD_{A}\eta_{pq}=D_{A}\eta_{pq}=\Phi_{Ap}{}^{r}\eta_{rq}+\Phi_{Aq}{}^{r}\eta_{pr}=0\,,\\
\cD_{A}\breta_{\brp\brq}=D_{A}\breta_{\brp\brq}=\brPhi_{A\brp}{}^{\brr}\breta_{\brr\brq}+\brPhi_{A\brq}{}^{\brr}\breta_{\brp\brr}=0\,,\\
\cD_{A}\Cp_{\alpha\beta}=D_{A}\Cp_{\alpha\beta}=\Phi_{A\alpha}{}^{\delta}\Cp_{\delta\beta}+\Phi_{A\beta}{}^{\delta}\Cp_{\alpha\delta}=0\,,\\
\cD_{A}\brCp_{\bralpha\brbeta}=D_{A}\brCp_{\bralpha\brbeta}=\brPhi_{A\bralpha}{}^{\brdelta}\brCp_{\brdelta\brbeta}+\brPhi_{A\brbeta}{}^{\brdelta}\brCp_{\bralpha\brdelta}=0\,,
\ea
\label{metrics}
\ee
and also with  all the gamma matrices,
\be
\ba{l}
\cD_{A}(\gamma^{p})^{\alpha}{}_{\beta}=D_{A}(\gamma^{p})^{\alpha}{}_{\beta}=\Phi_{A}{}^{p}{}_{q}(\gamma^{q})^{\alpha}{}_{\beta}
+\Phi_{A}{}^{\alpha}{}_{\delta}(\gamma^{p})^{\delta}{}_{\beta}-(\gamma^{p})^{\alpha}{}_{\delta}
\Phi_{A}{}^{\delta}{}_{\beta}=0\,,\\
\cD_{A}(\brgamma^{\brp})^{\bralpha}{}_{\brbeta}=D_{A}(\brgamma^{\brp})^{\bralpha}{}_{\brbeta}=\brPhi_{A}{}^{\brp}{}_{\brq}(\brgamma^{\brq})^{\bralpha}{}_{\brbeta}+\brPhi_{A}{}^{\bralpha}{}_{\brdelta}(\brgamma^{\brp})^{\brdelta}{}_{\brbeta}-(\brgamma^{\brp})^{\bralpha}{}_{\brdelta}
\brPhi_{A}{}^{\brdelta}{}_{\brbeta}=0\,.
\ea
\label{GG}
\ee
It follows then that the connections are all anti-symmetric,
\begin{eqnarray}
{}&&\quad\quad\quad\quad\Gamma_{ABC}=-\Gamma_{ACB}\,,\label{Gskew}\\
{}&&\Phi_{Apq}=-\Phi_{Aqp}\,,\quad\quad\brPhi_{A\brp\brq}=-\brPhi_{A\brq\brp}\,,\\
{}&&\Phi_{A\alpha\beta}=-\Phi_{A\beta\alpha}\,,\quad\quad\!\!\brPhi_{A\bralpha\brbeta}=-\brPhi_{A\brbeta\bralpha}\,,
\end{eqnarray}
and as usual,
\be
\ba{ll}
\Phi_{A}{}^{\alpha}{}_{\beta}=\quarter\Phi_{Apq}(\gamma^{pq})^{\alpha}{}_{\beta}\,,~~~~&~~~~
\brPhi_{A}{}^{\bralpha}{}_{\brbeta}=\quarter\brPhi_{A\brp\brq}(\brgamma^{\brp\brq})^{\bralpha}{}_{\brbeta}\,.
\ea
\ee
{}~\\

\indent Secondly,  the master semi-covariant derivative annihilates  all the NS-NS fields,  
\be
\ba{l}
\cD_{A}d=\na_{A}d:=-\half e^{2d}\na_{A}(e^{-2d})=\partial_{A}d+\half\Gamma^{B}{}_{BA}=0\,,\\
\cD_{A}V_{Bp}=\partial_{A}V_{Bp}+\Gamma_{AB}{}^{C}V_{Cp}+\Phi_{Ap}{}^{q}V_{Bq}=0\,,\\
\cD_{A}\brV_{B\brp}=\partial_{A}\brV_{B\brp}+\Gamma_{AB}{}^{C}\brV_{C\brp}+\brPhi_{A\brp}{}^{\brq}\brV_{B\brq}=0\,.
\ea
\label{VVd}
\ee
It follows  that
\be
\ba{ll}
\cD_{A}P_{BC}=\na_{A}P_{BC}=0\,,~~~~&~~~~\cD_{A}\brP_{BC}=\na_{A}\brP_{BC}=0\,,
\ea
\label{DPbrP}
\ee
and  the connections are  related to each other by
\be
\ba{l}
\Gamma_{ABC}=V_{B}{}^{p}D_{A}V_{Cp}+\brV_{B}{}^{\brp}D_{A}\brV_{C\brp}\,,\\
\Phi_{Apq}=V^{B}{}_{p}\na_{A}V_{Bq}\,,\\
\brPhi_{A\brp\brq}=\brV^{B}{}_{\brp}\na_{A}\brV_{B\brq}\,.
\ea
\label{c3}
\ee

The connections  assume  the following  most general forms~\cite{Jeon:2011sq},
\be
\ba{l}
\Gamma_{CAB}=\Gammao_{CAB}+\Delta_{Cpq}V_{A}{}^{p}V_{B}{}^{q}+\brDelta_{C\brp\brq}\brV_{A}{}^{\brp}\brV_{B}{}^{\brq}\,,\\
\Phi_{Apq}=\Phio_{Apq}+\Delta_{Apq}\,,\\
\brPhi_{A\brp\brq}=\brPhio_{A\brp\brq}+\brDelta_{A\brp\brq}\,.
\ea
\label{PhibrPhi}
\ee
Here, from  \cite{Jeon:2011cn},\footnote{For a recent rederivation of the connection (\ref{Gammao}) and related discussion, see  \cite{Hohm:2011si}.   }
\be
\ba{ll}
\Gammao_{CAB}=&2\left(P\partial_{C}P\brP\right)_{[AB]}
+2\left({{\brP}_{[A}{}^{D}{\brP}_{B]}{}^{E}}-{P_{[A}{}^{D}P_{B]}{}^{E}}\right)\partial_{D}P_{EC}\\
{}&-\textstyle{\frac{4}{D-1}}\left(\brP_{C[A}\brP_{B]}{}^{D}+P_{C[A}P_{B]}{}^{D}\right)\!\left(\partial_{D}d+(P\partial^{E}P\brP)_{[ED]}\right)\,,
\ea
\label{Gammao}
\ee
and,  in terms of this, with  the corresponding  derivative, $\DOo_{A}=\partial_{A}+\Gammao_{A}$,
\be
\ba{l}
\Phio_{Apq}=V^{B}{}_{p}\DOo_{A} V_{Bq}=V^{B}{}_{p}\partial_{A} V_{Bq}+\Gammao_{ABC}V^{B}{}_{p}V^{C}{}_{q}\,,\\
\brPhio_{A\brp\brq}=\brV^{B}{}_{\brp}\DOo_{A}\brV_{B\brq}=\brV^{B}{}_{\brp}\partial_{A}\brV_{B\brq}+\Gammao_{ABC}\brV^{B}{}_{\brp}\brV^{C}{}_{\brq}\,.
\ea
\label{Phio}
\ee
Further, the extra  pieces, $\Delta_{Apq}$ and $\brDelta_{A\brp\brq}$,  correspond to  the   `torsion'  of  SDFT, which must be  covariant and   satisfy~\cite{Jeon:2011sq}
\be
\ba{ll}
\Delta_{Apq}=-\Delta_{Aqp}\,,~~~~&~~~~\brDelta_{A\brp\brq}=-\brDelta_{A\brq\brp}\,,\\
\Delta_{Apq}V^{Ap}=0\,,~~~~&~~~~\brDelta_{A\brp\brq}\brV^{A\brp}=0\,.
\ea
\label{Torsioncon}
\ee
Otherwise they are arbitrary. That is to say,  with (\ref{Gammao}), (\ref{Phio}), (\ref{Torsioncon}),    the connections in (\ref{PhibrPhi}) provide   the most  general solution  to the constraints, (\ref{metrics}), (\ref{GG}) and (\ref{VVd}). Note that,  the latter two `traceless' conditions  in (\ref{Torsioncon}) are  necessary   in order  to maintain $\cD_{A}d=0$. As is the case in ordinary  supergravities,     the torsion can be constructed from the bi-spinorial  objects. We refer to  our earlier work~\cite{Jeon:2011sq}  for the torsions  in the case of ${\cN=1\,}$ ${D=10}\,$ SDFT. \\

The torsionless connection, $\Gammao_{ABC}$  (\ref{Gammao}),   further  obeys~\cite{Jeon:2011cn,Jeon:2011vx},
\be
\Gammao_{ABC}+\Gammao_{BCA}+\Gammao_{CAB}=0\,,
\label{torsionless}
\ee
and
\be
\ba{ll}
\cP_{CAB}{}^{DEF}\Gammao_{DEF}=0\,,~~~~&~~~~
\bcP_{CAB}{}^{DEF}\Gammao_{DEF}=0\,.
\ea
\label{RANK6used}
\ee
In fact,  the torsionless connection,  $\Gammao_{ABC}$ (\ref{Gammao}),  is the unique connection which satisfies these extra properties:  enforcing   (\ref{torsionless}) and (\ref{RANK6used}) on the generic connection, $\Gamma_{ABC}$ (\ref{PhibrPhi}),  would eliminate the torsion,  and  hence reduce  $\Gamma_{ABC}$ to $\Gammao_{ABC}$.   \\

The two  symmetric properties of the torsionless connection,  (\ref{Gskew})  and   (\ref{torsionless}),  enable us to replace  the ordinary derivatives in the definition
of the generalized Lie derivative (\ref{tcL})  by the torsion free   semi-covariant derivative,  $\DOo_{A}=\partial_{A}+\Gammao_{A}$~\cite{Jeon:2011cn,Jeon:2011vx}.   In a way,  the torsionless connection,  $\Gammao_{ABC}$ (\ref{Gammao}),  is the DFT analogy of  the   Christoffel connection in Riemannian geometry.  \\

It is also worth while to note, upon the section condition (\ref{constraint}), 
 \be
P_{I}{}^{A}\brP_{J}{}^{B}\Gamma^{C}{}_{AB}\partial_{C}\,\seceq\,0\,.
\label{USEFULv}
\ee
~\\

\indent The usual  field strengths of the three connections, 
\be
\ba{l}
R_{CDAB}=\partial_{A}\Gamma_{BCD}-\partial_{B}\Gamma_{ACD}+\Gamma_{AC}{}^{E}\Gamma_{BED}-\Gamma_{BC}{}^{E}\Gamma_{AED}\,,\\
F_{ABpq}=\partial_{A}\Phi_{Bpq}-\partial_{B}\Phi_{Apq}+\Phi_{Apr}\Phi_{B}{}^{r}{}_{q}-\Phi_{Bpr}\Phi_{A}{}^{r}{}_{q}\,,\\
\brF_{AB\brp\brq}=\partial_{A}\brPhi_{B\brp\brq}-\partial_{B}\brPhi_{A\brp\brq}+\brPhi_{A\brp\brr}\brPhi_{B}{}^{\brr}{}_{\brq}-\brPhi_{B\brp\brr}\brPhi_{A}{}^{\brr}{}_{\brq}\,,
\ea
\label{threeF}
\ee 
are, from   $\,[\cD_{A},\cD_{B}]V_{Cp}=0\,$  and   $\,[\cD_{A},\cD_{B}]\brV_{C\brp}=0$,~    related to each other through 
\be
R_{ABCD}=F_{CDpq}V_{A}{}^{p}V_{B}{}^{q}+\brF_{CD\brp\brq}\brV_{A}{}^{\brp}\brV_{B}{}^{\brq}\,.
\label{RFF}
\ee
It follows then  that
\be
\ba{ll}
R_{ABCD}=R_{[AB][CD]}\,,\quad&\quad P_{A}{}^{I}\brP_{B}{}^{J}R_{IJCD}=0\,.
\ea
\ee
However,  none of the field strengths in (\ref{threeF})  is   double-gauge covariant~\cite{Jeon:2011cn}.\\

In order to construct  DFT covariant   curvature tensors, it is necessary  to   first define~\cite{Jeon:2011cn},
\be
S_{ABCD}:=\half\left(R_{ABCD}+R_{CDAB}-\Gamma^{E}{}_{AB}\Gamma_{ECD}\right)\,.
\label{Sdef}
\ee
This rank-four field   satisfies,  precisely the same symmetric properties as the Riemann curvature,
\be
S_{ABCD}=\half\left(S_{[AB][CD]}+S_{[CD][AB]}\right)\,,
\label{Ssym}
\ee
as well as an additional  identity~\cite{Jeon:2011sq},
\be
P_{I}{}^{A}\brP_{J}{}^{B}P_{K}{}^{C}\brP_{L}{}^{D}S_{ABCD}\seceq0\,.
\label{Sid}
\ee
The latter holds up to the section condition  (\ref{constraint}) and further  implies   
\be
P^{AC}\brP^{BD}S_{ABCD}=-\half P^{AC}\brP^{BD}\Gamma^{E}{}_{AB}\Gamma_{ECD}\seceq0\,.
\ee
Another crucial property of $S_{ABCD}$ is that, under arbitrary variation of the connection, $\delta\Gamma_{ABC}$, it transforms as  
\be
\delta S_{ABCD}=\cD_{[A}\delta\Gamma_{B]CD}+\cD_{[C}\delta\Gamma_{D]AB}
-\textstyle{\frac{3}{2}}\Gamma_{[ABE]}\delta\Gamma^{E}{}_{CD}
-\textstyle{\frac{3}{2}}\Gamma_{[CDE]}\delta\Gamma^{E}{}_{AB}\,.
\label{Svar}
\ee
Further, from (\ref{PhibrPhi}),   if we write 
\be
\ba{ll}
\Gamma_{ABC}=\Gammao{}_{ABC}+\Lambda_{ABC}\,,~~~~&~~~
\Lambda_{ABC}=\Delta_{Apq}V_{B}{}^{p}V_{C}{}^{q}+\brDelta_{A\brp\brq}\brV_{B}{}^{\brp}\brV_{C}{}^{\brq}\,,
\ea
\ee
we get
\be
S_{ABCD}=S^{\scriptscriptstyle{0}}_{ABCD}+\cDo_{[A}\Lambda_{B]CD}+\cDo_{[C}\Lambda_{D]AB}+\Lambda_{D[A}{}^{E}\Lambda_{|C|B]E}+\Lambda_{B[C}{}^{E}\Lambda_{|A|D]E}-\half\Lambda^{E}{}_{AB}\Lambda_{ECD}\,.
\label{SABCD}
\ee
Consequently, with
\be
S_{AB}:=S^{C}{}_{ACB}\,,
\ee
we also obtain\footnote{Note that,  in contrast to  (\ref{SABCD}), we  have organized  the right hand side of the equality in (\ref{SAB}) as torsionful objects.}
\be
S^{\scriptscriptstyle{0}}_{AB}=S_{AB}+\cD_{C}\Lambda_{(AB)}{}^{C}+\half\Lambda_{CAD}\Lambda^{C}{}_{B}{}^{D}-\Lambda_{CAD}\Lambda^{D}{}_{B}{}^{C}\,.
\label{SAB}
\ee
Especially  for  the torsion free   case,  we have in addition to (\ref{Ssym}) and (\ref{Sid})~\cite{Jeon:2011cn}
\be
\ba{l}
P_{I}{}^{A}P_{J}{}^{B}\brP_{K}{}^{C}\brP_{L}{}^{D}\So_{ABCD}\seceq0\,,\\
P_{I}{}^{A}\brP_{J}{}^{C}(P^{BD}-\brP^{BD})\So_{ABCD}\seceq 0\,,\\
\So{}^{A}{}_{A}\seceq0\,,\\
(P^{AB}P^{CD}+\brP^{AB}\brP^{CD})\So_{ACBD}\seceq 0\,,\\
\So_{ABCD}+\So_{BCAD}+\So_{CABD}=0\quad\quad:\quad\quad\mbox{Bianchi~identitiy}\,,
\ea
\label{Bianchi}
\ee
and the relation~(\ref{Svar})  reduces to
\be
\delta \So_{ABCD}=\cDo_{[A}\delta\Gammao_{B]CD}+\cDo_{[C}\delta\Gammao_{D]AB}\,.
\ee
The variation of the torsionless connection ought to be  induced by the (arbitrary)  variations of the projections and the DFT-dilaton~\cite{Jeon:2011cn},
\be
\ba{ll}
\!{\delta\Gammao}_{CAB}=&2P_{[A}^{~D}\brP_{B]}^{~E}\DOo_{C}\delta P_{DE}+2(\brP_{[A}^{~D}\brP_{B]}^{~E}-P_{[A}^{~D}P_{B]}^{~E})\DOo_{D}\delta P_{EC}\\
{}&-\textstyle{\frac{4}{D-1}}(\brP_{C[A}\brP_{B]}^{~D}+P_{C[A}P_{B]}^{~D})(\partial_{D}\delta d+P_{E[G}\DOo{}^{G}\delta P^{E}_{~D]})\\
{}&-\Gammao_{FDE\,}\delta(\cP+\bcP)_{CAB}{}^{FDE}\,,
\ea
\label{useful0}
\ee
where the variations of the projections meet~\cite{Jeon:2010rw}
\be
\ba{lll}
\delta P=P\delta P\brP+\brP\delta P P\,,~~~~&~~~~
\delta\brP=P\delta\brP\brP+\brP\delta\brP P\,,~~~~&~~~~\delta P=-\delta \brP\,,
\ea
\ee
and  may be generated  by those  of the double-vielbein,
\be
\ba{ll}
\delta P_{AB}=V_{B}{}^{p}\delta V_{Ap}+V_{A}{}^{p}\delta V_{Bp}\,,~~~~&~~~~
\delta\brP_{AB}=\brV_{B}{}^{\brp}\delta\brV_{A\brp}+\brV_{A}{}^{\brp}\delta\brV_{B\brp}\,.
\ea
\label{useful1}
\ee
Further, the arbitrary variations of the double-vielbein satisfy 
\be
\ba{ll}
\delta V_{Ap}=\brP_{A}{}^{B}\delta V_{Bp}+V_{A}{}^{q}\delta V_{B[p}V^{B}{}_{q]}\,,~~~~&~~~~
\delta\brV_{A\brp}=P_{A}{}^{B}\delta\brV_{B\brp}+\brV_{A}{}^{\brq}\delta\brV_{B[\brp}\brV^{B}{}_{\brq]}\,,
\ea
\label{useful2}
\ee
and  the generic  torsionful  spin connections transform as
\be
\ba{l}
\delta\Phi_{Apq}=\cD_{A}(V^{B}{}_{p}\delta V_{Bq})+V^{B}{}_{p}V^{C}{}_{q}\delta\Gamma_{ABC}\,,\\
\delta\brPhi_{A\brp\brq}=\cD_{A}(\brV^{B}{}_{\brp}\delta \brV_{B\brq})+\brV^{B}{}_{\brp}\brV^{C}{}_{\brq}\delta\Gamma_{ABC}\,,
\ea
\label{useful3}
\ee
and  the gravitinos vary   as 
\be
\ba{l}
\delta\psi_{A}=\left(\delta\psi_{\brp}+\psi_{\brq}\delta\brV_{B}{}^{\brq}\brV^{B}{}_{\brp}\right)\brV_{A}{}^{\brp}
-\psi_{B}\delta V^{B}{}_{p}V_{A}{}^{p}\,,\\
\delta\psi^{\prime}_{A}=\left(\delta\psi^{\prime}_{p}+\psi^{\prime}_{q}\delta V_{B}{}^{q}V^{B}{}_{p}\right)V_{A}{}^{p}-\psi^{\prime}_{B}\delta \brV^{B}{}_{\brp}\brV_{A}{}^{\brp}\,.

\ea
\label{useful4}
\ee
~\\

\subsection{Projection-aided covariant derivatives and covariant curvatures\label{SECfullcovD}}

Under double-gauge  transformations,  the  connection and  the semi-covariant derivative transform as
\be
\ba{l}
(\delta_{X}{-\hcL_{X}})\Gamma_{CAB}\seceq 2\big[(\cP{+\bcP})_{CAB}{}^{FDE}-\delta_{C}^{~F}\delta_{A}^{~D}\delta_{B}^{~E}\big]\partial_{F}\partial_{[D}X_{E]}\,,\\
\dis{(\delta_{X}{-\hcL_{X}})\na_{C}T_{A_{1}\cdots A_{n}}\seceq
\sum_{i=1}^{n}2(\cP{+\bcP})_{CA_{i}}{}^{BFDE}
\partial_{F}\partial_{[D}X_{E]}T_{A_{1}\cdots A_{i-1} BA_{i+1}\cdots A_{n}}\,.}
\ea
\label{noncov}
\ee
Hence,  the semi-covariant derivative is not generically  double-gauge covariant.\footnote{Nevertheless exceptionally,    Eqs.(\ref{metrics}, \ref{GG}, \ref{VVd},  \ref{DPbrP}, \ref{torsionless},  \ref{RANK6used}) are double-gauge covariant. This  fact  is consistent with the uniqueness of the torsionless connection and the  covariant property   of the torsion.  }  We say, a tensor is double-gauge covariant if and only if its double-gauge transformation agrees  with  the   generalized Lie derivative,  \textit{i.e.~}`$\,\delta_{X}\seceq\hcL_{X}\,$'.  \\

Similarly, while the derivative $D_{A}=\partial_{A}+\Phi_{A}+\brPhi_{A}$ (\ref{nadd}) is   $\SpinD\times\oSpinD$   local Lorentz covariant, it is not double-gauge covariant, since
\be
\ba{l}
(\delta_{X}-\hcL_{X})\Phi_{Apq}\seceq2\cP_{ABC}{}^{DEF}\partial_{D}\partial_{[E}X_{F]}V^{B}{}_{p}V^{C}{}_{q}\,,\\
(\delta_{X}-\hcL_{X})\brPhi_{A\brp\brq}\seceq2\bcP_{ABC}{}^{DEF}\partial_{D}\partial_{[E}X_{F]}\brV^{B}{}_{\brp}\brV^{C}{}_{\brq}\,.
\ea
\ee
Also from 
\be
(\delta_{X}-\hcL_{X})\So_{ABCD}\seceq 2\cDo_{[A}\left((\cP{+\bcP})_{B][CD]}{}^{EFG}\partial_{E}\partial_{[F}X_{G]}\right)+2\cDo_{[C}\left((\cP{+\bcP})_{D][AB]}{}^{EFG}\partial_{E}\partial_{[F}X_{G]}\right)\,,
\ee
we see that $\So_{ABCD}$ is not double-gauge covariant as well.\\

However,    the characteristic feature  of the `semi-covariant'  derivative is that  ---as the name indicates---    combined with the projections,  it can  generate  various    fully covariant    quantities, with respect to double-gauge, $\SpinD\times\oSpinD$ double local Lorentz and $\ODD$ symmetries. \\

We write  down (projected) parts of spin connections which are  \textit{double-gauge covariant}~\cite{Jeon:2011cn,Jeon:2011vx},
\be
\ba{llllll}
\brP_{A}{}^{B}\Phi_{Bpq}\,,~&~P_{A}{}^{B}\brPhi_{B\brp\brq}\,,~&~\Phi_{A[pq}V^{A}{}_{r]}\,,~&~
\brPhi_{A[\brp\brq}\brV^{A}{}_{\brr]}\,,~&~\Phi_{Apq}V^{Ap}\,,~&~\brPhi_{A\brp\brq}\brV^{A\brp}\,.
\ea
\label{covPhi}
\ee 
~\\
From these,\,  \underline{\textit{fully covariant quantities}}\,  follow.\\

\begin{itemize}
\item \textbf{Covariant derivatives for $\ODD$ tensors}~\cite{Jeon:2011cn}:
\be
\ba{c}
P_{C}{}^{D}{\brP}_{A_{1}}{}^{B_{1}}{\brP}_{A_{2}}{}^{B_{2}}\cdots{\brP}_{A_{n}}{}^{B_{n}}
\DO_{D}T_{B_{1}B_{2}\cdots B_{n}}\,,\\
{\brP}_{C}{}^{D}P_{A_{1}}{}^{B_{1}}P_{A_{2}}{}^{B_{2}}\cdots P_{A_{n}}{}^{B_{n}}
\DO_{D}T_{B_{1}B_{2}\cdots B_{n}}\,,\\
P^{AB}{\brP}_{C_{1}}{}^{D_{1}}{\brP}_{C_{2}}{}^{D_{2}}\cdots{\brP}_{C_{n}}{}^{D_{n}}\DO_{A}T_{BD_{1}D_{2}\cdots D_{n}}\,,\\
\brP^{AB}{P}_{C_{1}}{}^{D_{1}}{P}_{C_{2}}{}^{D_{2}}\cdots{P}_{C_{n}}{}^{D_{n}}\DO_{A}T_{BD_{1}D_{2}\cdots D_{n}}\,,\\
P^{AB}{\brP}_{C_{1}}{}^{D_{1}}{\brP}_{C_{2}}{}^{D_{2}}\cdots{\brP}_{C_{n}}{}^{D_{n}}
\DO_{A}\DO_{B}T_{D_{1}D_{2}\cdots D_{n}}\,,\\
{\brP}^{AB}P_{C_{1}}{}^{D_{1}}P_{C_{2}}{}^{D_{2}}\cdots P_{C_{n}}{}^{D_{n}}
\DO_{A}\DO_{B}T_{D_{1}D_{2}\cdots D_{n}}\,.
\ea
\label{covT}
\ee
~\\

\item \textbf{Covariant derivatives for $\SpinD\times\oSpinD$ tensors}:
\be
\ba{ll}
\cD_{p}T_{\brq_{1}\brq_{2}\cdots\brq_{n}}\,,~~~~&~~~~
\cD_{\brp}T_{q_{1}q_{2}\cdots q_{n}}\,,\\
\cD^{p}T_{p\brq_{1}\brq_{2}\cdots\brq_{n}}\,,~~~~&~~~~
\cD^{\brp}T_{\brp q_{1}q_{2}\cdots q_{n}}\,,\\
\cD_{p}\cD^{p}T_{\brq_{1}\brq_{2}\cdots\brq_{n}}\,,~~~~&~~~~
\cD_{\brp}\cD^{\brp}T_{q_{1}q_{2}\cdots q_{n}}\,,
\ea
\label{covT2}
\ee
where we set
\be
\ba{ll}
\cD_{p}:=V^{A}{}_{p}\cD_{A}\,,~~~~&~~~~\cD_{\brp}:=\brV^{A}{}_{\brp}\cD_{A}\,.
\ea
\label{Dpbrp}
\ee
These are simply the pull-back of the chiral and anti-chiral $\ODD$ vector indices in (\ref{covT}) to the $\SpinD$ and $\oSpinD$ vector indices using the double-vielbeins. 
~\\

\item \textbf{Covariant Dirac operators for fermions, $\rho^{\alpha}$, $\,\psi^{\alpha}_{\brp}$,  $\,\rhop^{\bralpha}$, $\,\psi^{\prime\bralpha}_{p}$\,}~\cite{Jeon:2011vx,Jeon:2011sq}:
\be
\ba{ll}
\gamma^{p}\cD_{p}\rho=\gamma^{A}\cD_{A}\rho\,,~~~~&~~~~\gamma^{p}\cD_{p}\psi_{\brp}=\gamma^{A}\cD_{A}\psi_{\brp}\,,\\
\cD_{\brp}\rho\,,~~~~&~~~~\cD_{\brp}\psi^{\brp}=\cD_{A}\psi^{A}\,,\\
\multicolumn{2}{c}{\brpsi^{A}\gamma_{p}(\cD_{A}\psi_{\brq}-\half\cD_{\brq}\psi_{A})\,,~~~}
\ea
\label{covDirac}
\ee
and\footnote{Writing explicitly,
\[
\ba{ll}
\cD_{A}\psi_{\brp}=(\partial_{A}+\quarter\Phi_{Apq}\gamma^{pq})\psi_{\brp}
+\brPhi_{A\brp}{}^{\brq}\psi_{\brq}\,,~~~&~~~
\cD_{A}\psi_{B}=(\partial_{A}+\quarter\Phi_{Apq}\gamma^{pq})\psi_{B}+\Gamma_{AB}{}^{C}\psi_{C}\,,\\
\cD_{A}\psi^{\prime}_{p}=(\partial_{A}+\quarter\brPhi_{A\brp\brq}\brgamma^{\brp\brq})\psi^{\prime}_{p}
+\Phi_{Ap}{}^{q}\psi^{\prime}_{q}\,,~~~&~~~
\cD_{A}\psi^{\prime}_{B}=(\partial_{A}+\quarter\brPhi_{A\brp\brq}\brgamma^{\brp\brq})\psi^{\prime}_{B}+\Gamma_{AB}{}^{C}\psi^{\prime}_{C}\,.
\ea
\]}
\be
\ba{ll}
\brgamma^{\brp}\cD_{\brp}\rhop=\brgamma^{A}\cD_{A}\rhop\,,~~~~&~~~~\brgamma^{\brp}\cD_{\brp}\psi^{\prime}_{p}=\brgamma^{A}\cD_{A}\psi^{\prime}_{p}\,,\\
\cD_{p}\rhop\,,~~~~&~~~~\cD_{p}\psip^{p}=\cD_{A}\psip^{A}\,,\\
\multicolumn{2}{c}{\brpsi^{\prime A}\brgamma_{\brp}(\cD_{A}\psi^{\prime}_{q}-\half\cD_{q}\psi^{\prime}_{A})\,.~~~}
\ea
\label{covDiracp}
\ee
~\\

\item  \textbf{Covariant derivatives for $\SpinD\times\oSpinD$ bi-fundamental spinor, $\cT^{\alpha}{}_{\brbeta}$\,}: 
\be
\ba{ll}
\gamma^{A}\cD_{A}\cT\,,~~~~~&~~~~~~\cD_{A}\cT\brgamma^{A}\,.
\ea
\label{NewResult}
\ee
These  are new results  we report in this paper.  Combining the  two, we further define
\be
\ba{l}
\cD_{+}\cT:=\gamma^{A}\cD_{A}\cT+\gamma^{(D+1)}\cD_{A}\cT\brgamma^{A}\,,\\
\cD_{-}\cT:=\gamma^{A}\cD_{A}\cT-\gamma^{(D+1)}\cD_{A}\cT\brgamma^{A}\,.
\ea
\label{Dpm}
\ee
As shown in section~\ref{SECnil}, for the torsion free  case, the corresponding operators   are \textit{nilpotent},  up to the  section condition  (\ref{constraint}),
\be
\ba{ll}
(\cDo_{+})^{2}\cT\seceq 0\,,~~~~~&~~~~~(\cDo_{-})^{2}\cT\seceq0\,,
\ea
\label{nilpotent} 
\ee 
and hence, they  define cohomology.   \\

It is worth while  to note
\be
\ba{ll}
\cD_{\pm}(\gamma^{(D+1)}\cT)=-\gamma^{(D+1)}\cD_{\mp}\cT\,,~~~~&~~~~
\cD_{\pm}(\cT\brgamma^{(D+1)})=(\cD_{\mp}\cT)\brgamma^{(D+1)}\,.
\ea
\label{DgD}
\ee
~\\

\item  \textbf{Covariant curvatures}~\cite{Jeon:2011cn,Jeon:2011sq}:\\
Scalar curvatures, 
\be
\ba{llll}
P^{AB}P^{CD}S_{ACBD}\,,~~~~&~~~~~\brP^{AB}\brP^{CD}S_{ACBD}\,,~~~&~~~~P^{AB}S_{AB}\,,~~~&~~~~
\brP^{AB}S_{AB}\,,
\ea
\ee
and a rank-two curvature, 
\be
S_{p\brq}+\half\cD_{\brr}\brDelta_{p\brq}{}^{\brr}+\half\cD_{r}\Delta_{\brq p}{}^{r}\,,
\label{Spbrq}
\ee
 where  we set
 \be
 S_{p\brq}:=V^{A}{}_{p}\brV^{B}{}_{\brq}S_{AB}\,.
 \label{Spbrp}
 \ee
 We emphasize that ---while  $\So_{p\brq}$ is covariant   for the torsionless connection---  if nontrivial  torsion is present,  
 $S_{p\brq}$ alone is not covariant:  the full expression in Eq.(\ref{Spbrq}) is called upon as for a covariant quantity.\footnote{For example, see  the equations of motion in  $\cN=1$ SDFT~\cite{Jeon:2011sq}.}    \\

 It is worth while to note, up to the section condition (\ref{constraint}), 
 \be
 \ba{l}
 P^{AB}P^{CD}S_{ACBD}\seceq P^{AB}S_{AB}\,,\\
 \brP^{AB}\brP^{CD}S_{ACBD}\seceq\brP^{AB}S_{AB}\,.
 \ea
 \label{Sidequiv}
 \ee
Further, in the torsion free case,  all the scalar  curvatures are equivalent as
\be
P^{AB}P^{CD}\So_{ACBD}\seceq P^{AB}\So_{AB}\seceq-\brP^{AB}\brP^{CD}\So_{ACBD}\seceq-\brP^{AB}\So_{AB}\,.
\ee
While any  of them may  constitute the  DFT Lagrangian for the  NS-NS sector~\cite{Jeon:2011cn,Jeon:2011sq}, only the following combination  allows for the  1.5 formalism to work in   supersymmetric double field theory~\cite{Jeon:2011sq},
\be
\cL_{\NS}=\textstyle{\frac{1}{8}}e^{-2d}(P^{AB}P^{CD}-\brP^{AB}\brP^{CD})S_{ACBD}\,.
\label{NSNSL}
\ee
\end{itemize}
~\\


\subsection{Reduction to Riemannian geometry in  $D$ dimension\label{SUBSECREDAPP}} 
Assuming that the upper half blocks are non-degenerate,  the  double-vielbein  satisfying the defining properties~(\ref{defV}) takes the following most general form~\cite{Jeon:2011cn},
\be
\ba{ll}
V_{Ap}=\textstyle{\frac{1}{\sqrt{2}}}{{\left(\ba{c} (e^{-1})_{p}{}^{\mu}\\(B+e)_{\nu p}\ea\right)}}\,,~~~~
&~~~~\brV_{A{\brp}}=\textstyle{\frac{1}{\sqrt{2}}}\left(\ba{c} (\bre^{-1})_{\brp}{}^{\mu}\\(B+\bre)_{\nu{\brp}}\ea\right)\,.
\ea
\label{Vform1}
\ee
Here  $e_{\mu}{}^{p}$ and  $\bre_{\nu}{}^{{\brp}}$ are two copies of  the $D$-dimensional   vielbein  corresponding  to  the same  spacetime metric,   
\be
e_{\mu}{}^{p}e_{\nu}{}^{q}\eta_{pq}=-\bre_{\mu}{}^{{\brp}}\bre_{\nu}{}^{\brq}\breta_{\brp\brq}=g_{\mu\nu}\,,
\ee
and $B_{\mu\nu}$ corresponds to the Kalb-Ramond two-form gauge field.  
We also set  in (\ref{Vform1}),
\be
\ba{ll}
B_{\mu p}=B_{\mu\nu}(e^{-1})_{p}{}^{\nu}\,,~~~~&~~~~B_{\mu\brp}=B_{\mu\nu}(\bre^{-1})_{{\brp}}{}^{\nu}\,.
\ea
\ee
It is worth while to note that,  $(\bre^{-1}e)_{\brp}{}^{p}$ and $(e^{-1}\bre)_{p}{}^{\brp}$ are local Lorentz transformations, such that
\be
\ba{ll}
(\bre^{-1}e)_{\brp}{}^{p}(\bre^{-1}e)_{\brq}{}^{q}\eta_{pq}=-\breta_{\brp\brq}\,,~~~~&~~~~
(e^{-1}\bre)_{p}{}^{\brp}(e^{-1}\bre)_{q}{}^{\brq}\breta_{\brp\brq}=-\eta_{pq}\,.
\ea
\label{ebreeta}
\ee
~\\

Instead of   (\ref{Vform1}), we may   choose the following alternative parametrization, 
\be
\ba{ll}
V_{A}{}^{p}=\textstyle{\frac{1}{\sqrt{2}}}{{\left(\ba{c}(\beta+\tilde{e})^{\mu p}\\
(\tilde{e}^{-1})^{p}{}_{\nu}\ea\right)\,,}}~~~~
&~~~~\brV_{A}{}^{{\brp}}=\textstyle{\frac{1}{\sqrt{2}}}{{\left(\ba{c}({\beta+\bar{\tilde{e}}})^{\mu p}\\
(\bar{\tilde{e}}^{-1})^{p}{}_{\nu}\ea\right)\,,}}
\ea
\label{Vform2}
\ee
with
\be
\ba{ll}
\beta^{\mu p}=\beta^{\mu\nu}(\tilde{e}^{-1})^{p}{}_{\nu}\,,~~~~&~~~~
\beta^{\mu \brp}=\beta^{\mu\nu}(\bar{\tilde{e}}^{-1})^{p}{}_{\nu}\,.
\ea
\ee
Physically, $\tilde{e}^{\mu}{}_{p}$ and  $\bar{\tilde{e}}{}^{\mu}{}_{\brp}$  correspond to a pair of  T-dual vielbeins, such that  both give rise to the winding mode spacetime  metric,
\be
\tilde{e}{}^{\mu}{}_{p}\tilde{e}{}^{\nu}{}_{q}\eta^{pq}=-
\bar{\tilde{e}}{}^{\mu}{}_{\brp}\bar{\tilde{e}}{}^{\nu}{}_{\brq}\eta^{\brp\brq}
=(g-Bg^{-1}B)^{-1\, \mu\nu}\,.
\ee
Note that, in the winding  mode sector, the $D$-dimensional curved spacetime indices are all upside-down,  such as      
$\tilde{x}_{\mu}$,   $\tilde{e}^{\mu}{}_{p}$,  $\bar{\tilde{e}}{}^{\mu}{}_{\brp}$, $\beta^{\mu\nu}$  (\textit{cf.}  $x^{\mu}$, $e_{\mu}{}^{p}$, $\bre_{\mu}{}^{\brp}$, $B_{\mu\nu}$). \\

In connection to the  section condition,  $\partial^{A}\partial_{A}\seceq 0$ (\ref{constraint}), the former  parametrization   (\ref{Vform1}) matches well with the choice, $\frac{\partial~}{\partial \tilde{x}_{\mu}}\seceq 0$, while the latter is natural when $\frac{\partial~}{\partial{x}^{\mu}}\seceq 0$. Yet if we consider  dimensional reductions  from $D$ to lower dimensions, there is no longer preferred parametrization. For related discussions  we refer to  \cite{Andriot:2011uh,Andriot:2012wx,Andriot:2012an,Aldazabal:2011nj,Geissbuhler:2011mx,GeissbuhlerThesis,Grana:2012rr}.  \\

Henceforth  we restrict ourselves to  the parametrization (\ref{Vform1}) and to the section choice,\footnote{This restriction  reduces our $\ODD$ covariant stringy differential geometry  to the generalized geometry.} 
\be
 \frac{\partial\,~}{\partial{\tx}_{\mu}}\seceq 0\,.
 \label{sectionchoice}
\ee
~\\

In analogy to the DFT master semi-covariant derivative, $\cD_{A}$~(\ref{cDA}),  we  consider a   genuinely  $D$-dimensional master  derivative~\cite{Jeon:2011cn},
\be
D_{\mu}=\trd_{\mu}+\omega_{\mu}+\bromega_{\mu}\,,
\label{DD}
\ee
which is  covariant   with respect to the  $D$-dimensional diffeomorphism and the pair of local Lorentz groups, $\SpinD\times\oSpinD$, as 
 $\trd_{\mu}$ denotes the standard $D$-dimensional   diffeomorphism  covariant derivative,  while  $\omega_{\mu}$ and $\bromega_{\mu}$ correspond to the spin connections of the local Lorentz groups, $\SpinD$ and $\oSpinD$ respectively.  Yet, it is not $\ODD$ covariant. \\

It satisfies 
\be
\ba{lll}
{D_{\lambda}g_{\mu\nu}=\trd_{\lambda}g_{\mu\nu}=0}\,,~~~~&~~~~{D_{\mu}e_{\nu m}=0}\,,~~~~&~~~~{D_{\mu}\bre_{\nu \brn}=0}\,,
\ea
\ee 
and hence,  as in  (\ref{c3}), the $D$-dimensional  spin connections are determined by
\be
\ba{ll}
\omega_{\mu mn}{=(e^{-1})_{m}{}^{\nu}\trd_{\mu}e_{\nu n}}\,,~~~~&~~~~\bromega_{\mu \brm\brn}{=(\bre^{-1})_{\brm}{}^{\nu}\trd_{\mu}\bre_{\nu \brn}}\,.  
\ea
\label{omega2}
\ee
 For the    diffeomorphism  covariant derivative,  $\trd_{\mu}$,   we assume   the  torsionless  Christoffel connection,
\be
\Big\{{}_{\mu}{}^{\lambda}{}_{\nu}\Big\}=\half g^{\lambda\kappa}\left(
\partial_{\mu}g_{\kappa\nu}+\partial_{\nu}g_{\mu\kappa}-\partial_{\kappa}g_{\mu\nu}\right)\,.
\ee
In terms of the parametrization of the double-vielbein (\ref{Vform1}),   the projection-aided   covariant  spin connections in Eq.(\ref{covPhi}) read  explicitly for the torsionless connection,
\be
\ba{ll}
\sqrt{2} \bar{V}^{A}{}_{\bar{p}} \Phio_{Aqr}\seceq  \bar{e}^{\mu}{}_{\bar{p}} \omega_{\mu qr} + \half H_{\brp qr}\,,~~~~&~~~~\sqrt{2} V^{A}{}_{p} \brPhio_{A \bar{q} \bar{r}} \seceq e^{\mu}{}_{p} \bar{\omega}_{\mu \bar{q} \bar{r}} + \half H_{p\brq\brr}\,,\\
\sqrt{2} \Phio_{A [pq}V^{A}{}_{r]} \seceq \omega_{\mu [pq} e^{\mu}{}_{r]} + \frac{1}{6} H_{pqr}\,,~~~~&~~~~
 \sqrt{2}  \brPhio_{A [\bar{p} \bar{q}} \bar{V}^{A}{}_{\bar{r}]} \seceq \bar{\omega}_{\mu [\bar{p} \bar{q}} \bar{e}^{\mu}{}_{\bar{r}]} + \frac{1}{6} H_{\brp\brq\brr}\,,\\
 \sqrt{2} V^{A p} \Phio_{Apq} \seceq e^{\mu p} \omega_{\mu pq} -2 \partial_{q} \phi\,,~~~~&~~~~
\sqrt{2} \bar{V}^{A \bar{p}}  \brPhio_{A \bar{p} \bar{q}} \seceq \bar{e}^{\mu \bar{p}} \bar{\omega}_{\mu \bar{p} \bar{q}} - 2 \partial_{\bar{q}} \phi\,.
\ea
\ee
Clearly, these are diffeomorphism and $B$-field gauge symmetry covariant, and hence, as asserted, double-gauge covariant.    Using the results, we may  express  all the fully covariant derivatives  in section \ref{SECfullcovD} in terms of the usual   
$D$-dimensional Riemannian terminology~\cite{Jeon:2011cn,Jeon:2011vx}. For example, for the fermions, $\rho^{\alpha}$, $\psi^{\alpha}_{\brp}$, we get
\be
\ba{l}
\sqrt{2} \gamma^{A}\cDo_{A}\rho\seceq \gamma^{m} \left( \partial_{m} \rho + \frac{1}{4} \omega_{m n p} \gamma^{n p} \rho + \frac{1}{24} H_{m n p} \gamma^{n p} \rho - \partial_{m} \phi \rho \right) \,,\\
\sqrt{2}\gamma^{A}\cDo_{A}\psi_{\brp}\seceq \gamma^{m} \left( \partial_{m} \psi_{\brp} + \frac{1}{4} \omega_{m n p} \gamma^{n p} \psi_{\brp} + \bar{\omega}_{m \brp\brq} \psi^{\bar{q}} + \frac{1}{24} H_{m n p} \gamma^{n p} \psi_{\brp} + \half H_{m \brp \brq} \psi^{\brq} - \partial_{m} \phi \psi_{\brp}
\right)\,,\\
\sqrt{2}\brV^{A}{}_{\brp}\cDo_{A}\rho\seceq \partial_{\brp} \rho + \frac{1}{4} \omega_{\brp q r} \gamma^{q r} \rho + \frac{1}{8} H_{\brp q r} \gamma^{q r} \rho \,,\\
\sqrt{2}\cDo_{A}\psi^{A}\seceq \partial^{\brp} \psi_{\brp} + \frac{1}{4} \omega_{\brp qr} \gamma^{q r} \psi^{\brp} +\bar{\omega}^{\brp}{}_{\brp \brq} \psi^{\brq} +  \frac{1}{8} H_{\brp q r} \gamma^{q r} \psi^{\brp} - 2 \partial_{\brp} \phi \psi^{\brp} \,,
\ea
\ee
and for the other fermions, $\rho^{\prime\bralpha}$, $\psi^{\prime\bralpha}_{p}$, which are in the opposite $\SpinD\times\oSpinD$ representations,  we have
\be
\ba{l}
\sqrt{2} \brgamma^{A}\cDo_{A}\rhop\seceq \brgamma^{\brm} \left( \partial_{\brm} \rho + \frac{1}{4} \bromega_{\brm \brn \brp} \brgamma^{\brn\brp} \rhop + \frac{1}{24} H_{\brm\brn\brp} \brgamma^{\brn\brp}\rhop - \partial_{\brm} \phi \rhop \right) \,,
\\
\sqrt{2}\brgamma^{A}\cDo_{A}\psi^{\prime}_{p}\seceq \brgamma^{\brm} \left( \partial_{\brm} \psi^{\prime}_{p} + \frac{1}{4} \bromega_{\brm\brn\brp} \brgamma^{\brn\brp} \psi^{\prime}_{p} + {\omega}_{\brm pq} \psip{}^{q} + \frac{1}{24} H_{\brm\brn\brp} \brgamma^{\brn\brp} \psi^{\prime}_{p} + \half H_{\brm p q} \psip{}^{q} - \partial_{\brm} \phi \psi^{\prime}_{p}\right)\,,\\
\sqrt{2}V^{A}{}_{p}\cDo_{A}\rhop\seceq \partial_{p} \rhop + \frac{1}{4} \bromega_{p \brq \brr} \brgamma^{\brq\brr} \rhop + \frac{1}{8} H_{p\brq\brr} \brgamma^{\brq\brr} \rhop \,,\\
\sqrt{2}\cDo_{A}\psip^{A}\seceq \partial^{p} \psi^{\prime}_{p} + \frac{1}{4} \bromega_{p\brq\brr} \brgamma^{\brq\brr} \psip^{p} +{\omega}^{p}{}_{pq} \psip^{q} +  \frac{1}{8} H_{p \brq\brr} \gamma^{\brq\brr} \psip^{p} - 2 \partial_{p} \phi \psip^{p} \,.
\ea
\ee
Here, for simplicity,  we set  $\partial_{p}=(e^{-1})_{p}{}^{\mu}\partial_{\mu}$, $\partial_{\brp}=(\bre^{-1})_{\brp}{}^{\mu}\partial_{\mu}$, \textit{etc.}\\~\\


\end{document}